%% file: main.tex
\documentclass[10pt, conference, letterpaper]{IEEEtran}

\usepackage{listings}

\usepackage{blindtext}
\usepackage{amsfonts}
\usepackage{amsmath, amsthm, amssymb}
\usepackage{times}
\usepackage{verbatim}
\usepackage{array}
\usepackage{mathrsfs}
\usepackage{enumitem}
\usepackage{textcomp}

\usepackage[ruled, lined, linesnumbered]{algorithm2e}
\SetAlFnt{\small}

\SetCommentSty{mycommfont}

\usepackage{amsfonts}
\usepackage{bbm}
\usepackage{soul}
\usepackage{color}
\usepackage{xcolor}
\usepackage{graphics}
\usepackage{graphicx,epsfig}
\usepackage{lipsum,graphicx,subcaption}
\graphicspath{{figures/}}
\usepackage{caption}
\usepackage{subcaption}

\usepackage{cleveref}
\usepackage{wrapfig}
\usepackage{xurl}

\usepackage{mathtools}

\usepackage{comment}
\usepackage{multirow}
\usepackage{tablefootnote}
\usepackage{footnote}
\usepackage{tablefootnote}

\renewcommand\IEEEkeywordsname{Keywords}

\newcommand{\note}[1]{{\color{red}{\bf #1}}} 
\newcommand{\greg}[1]{{\color{teal}{\bf #1}}} 
 
\begin{document}
\setlength{\abovedisplayskip}{5pt}
\setlength{\belowdisplayskip}{5pt}
\def\eg{\mbox{\em e.g.}, }


\title{Collaboration with Cellular Networks\\ for RFI Cancellation at Radio Telescope}

\author{\IEEEauthorblockN{Shuvam Chakraborty\IEEEauthorrefmark{1},
Gregory Hellbourg\IEEEauthorrefmark{2}, Maqsood Careem\IEEEauthorrefmark{1}, 
Dola Saha\IEEEauthorrefmark{1} and 
Aveek Dutta\IEEEauthorrefmark{1}}\\
\IEEEauthorblockA{\IEEEauthorrefmark{1}Department of Electrical and Computer Engineering, 
University at Albany, SUNY\\
\IEEEauthorrefmark{2}Department of Astronomy, 
California Institute of Technology\\
Email: \IEEEauthorrefmark{1}schakraborty@albany.edu,
\IEEEauthorrefmark{2}ghellbourg@astro.caltech.edu,
\IEEEauthorrefmark{1}mabdulcareem@albany.edu,\\
\IEEEauthorrefmark{1}dsaha@albany.edu, \IEEEauthorrefmark{1}adutta@albany.edu}}


\maketitle

\input abstract.tex


\input{intro.tex}

\input{prelim.tex}
\input{system.tex}
\input{rqf.tex}
\input{results.tex}

\vspace{-5pt}
\input{related.tex}

\vspace{-5pt}

\input{conclusion.tex}

\vspace{-5pt}

\section*{Acknowledgement}
\vspace{-5pt}
This work is funded by the National Science Foundation SWIFT Program (Award Number - 2128581).
\vspace{-5pt}

\bibliographystyle{IEEEtran}

\bibliography{references, fromSWIFT/dola_references_edited, fromSWIFT/greg_refs, fromSWIFT/ref}

\end{document}

%% file: abstract.tex
\begin{abstract}

The growing need for electromagnetic spectrum to support the next generation (xG) communication networks increasingly generate unwanted radio frequency interference (RFI) in protected bands for radio astronomy.
RFI is commonly mitigated at the Radio Telescope without any active collaboration with the interfering sources.
In this work, we provide a method of signal characterization and its use in subsequent cancellation, that uses Eigenspaces derived from the telescope and the transmitter signals. This is different from conventional time-frequency domain analysis, which is limited to fixed characterizations (\eg complex exponential in Fourier methods) that cannot adapt to the changing statistics (\eg autocorrelation) of the RFI, typically observed in communication systems. We have presented effectiveness of this collaborative method using real-world astronomical signals and practical simulated LTE signals (downlink and uplink) as source of RFI along with propagation conditions based on preset benchmarks and standards. Through our analysis and simulation using these signals, we are able to remove 89.04\% of the RFI from cellular networks, which reduces excision at the Telescope and capable of significantly improving throughput as corrupted time frequency bins data becomes usable.
\end{abstract}



%% file: intro.tex
\section{Introduction}
\label{sec:intro}

Radio Astronomy is a discovery-based science, which has revolutionized our understanding of the Universe through scientific observations across the electromagnetic (EM) spectrum. However, only 1-2\% of the spectrum is allocated for science below 50 GHz where almost all of the commercial radio communication occurs.
It is increasingly essential to use spectrum other than the current allocation for astronomical observations for two main reasons: 1) red-shifting of spectral lines due to the expanding Universe and 2) broad bandwidth radio continuum observations can increase the signal-to-noise ratio of weak radio sources. 
Radio telescopes are generally located in geographically isolated areas to avoid radio frequency interference (RFI) from human generated electromagnetic waves. However, no matter how remote, all radio astronomy sites across the world are vulnerable to growing RFI from terrestrial networks used to extend coverage to increasing human population~\cite{quiet_skies_20, nap21729_15}. 
Furthermore, proliferation of next generation (xG) communication networks~\cite{fcc_rural,rau2019rfi} increasingly generates RFI, even in bands that are protected for radio astronomy due to out-of-band emissions and intermodulation products.
At the same time, technological advances enable the development of wideband and low system temperature receivers, resulting in dramatic improvement to the sensitivity of modern radio telescopes to faint signals of astronomical origin~\cite{8742126,9440921,8879003}.
Generally, communication system designers strive to reduce noise from artificially generated signal, whereas radio astronomy focuses on removing communication signals from the astronomical signal. 
This seemingly opposing requirement is pushing the two communities farther away.
Both are equally essential and in essence are designed to overcome a common bottleneck: \textit{RFI}. So, we present a collaborative framework to address this problem, by aggregating concise, yet accurate signal characterization from the RFI source, 
and \textit{cancel} it at the telescope.


Current RFI mitigation techniques use statistical signal analysis to detect RFI and 
discard
the associated time and frequency bins from the collected data - defined as excision.
The excision of corrupted data may at best reduce the sensitivity of the telescope, and at worse remove the astronomical signal of interest or lead to erroneous result. In this work, we focus on the sub-6GHz bands
as it is among the busiest spectral windows, heavily exploited by commercial wireless applications, while offering a unique opportunity to observe astronomical emissions like continuum synchrotron emissions (e.g. pulsars show strongest emissions below 600 MHz), or at-rest and redshifted spectral lines like Neutral Atomic Hydrogen or Hydroxyl~\cite{national2007handbook}.
As a source of RFI, we have identified 4G/5G cellular service, which transmits at a high power, covers a large bandwidth ($\sim$ 2 GHz), most abundant and can propagate large distances. However, the proposed method is applicable to remove a variety of RFI in other frequency bands as well. 

\begin{figure}
\centering
        \begin{subfigure}[t]{0.24\textwidth}
           	\includegraphics[width=\linewidth]{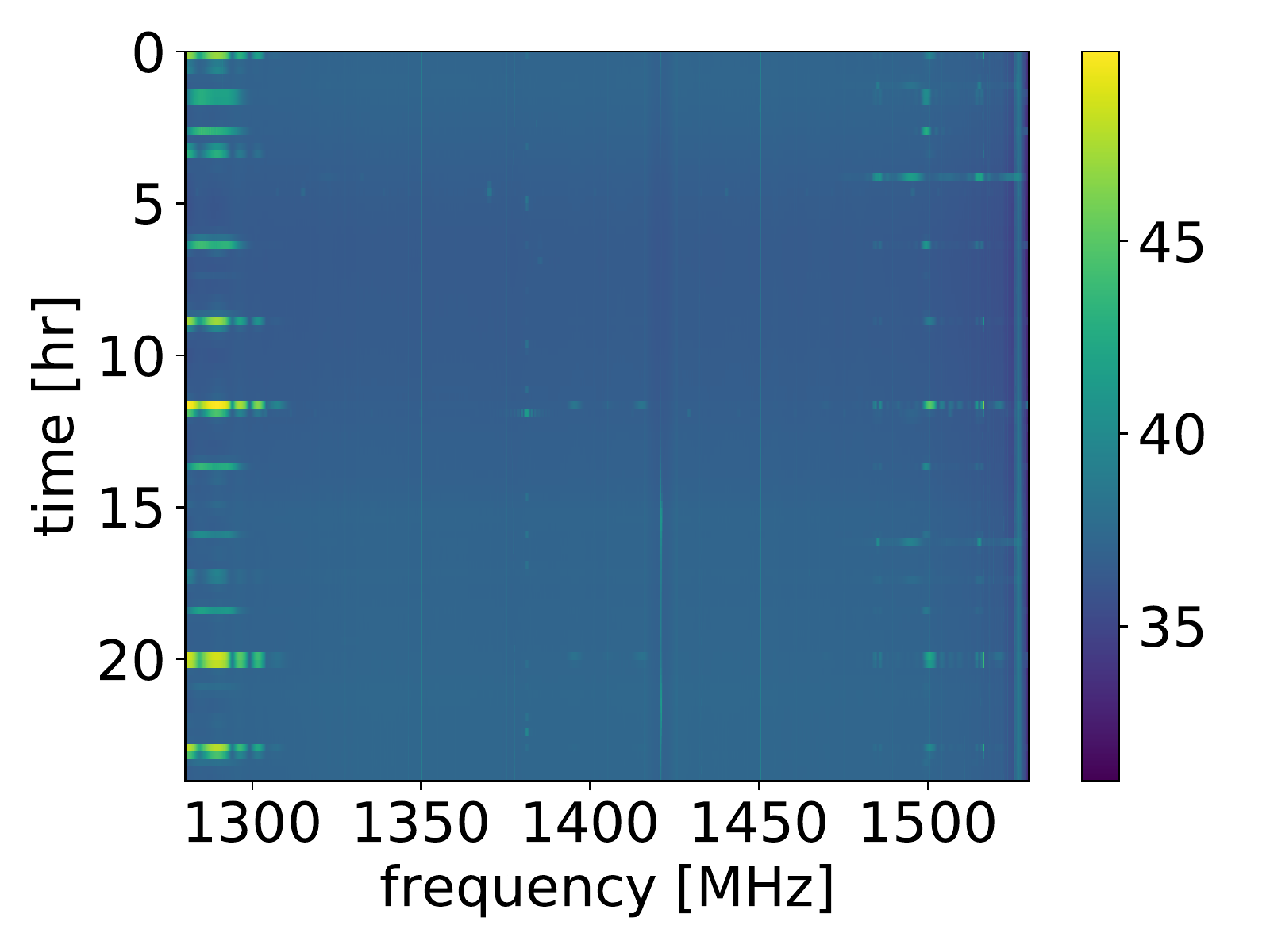}
    		\caption{DSA-110 spectrogram over 24 hours. The data are affected with various types of RFI, including telemetry and radar systems.}
    		\label{fig:contaminated}       
        \end{subfigure}
        \begin{subfigure}[t]{0.24\textwidth}
            \includegraphics[width=\linewidth]{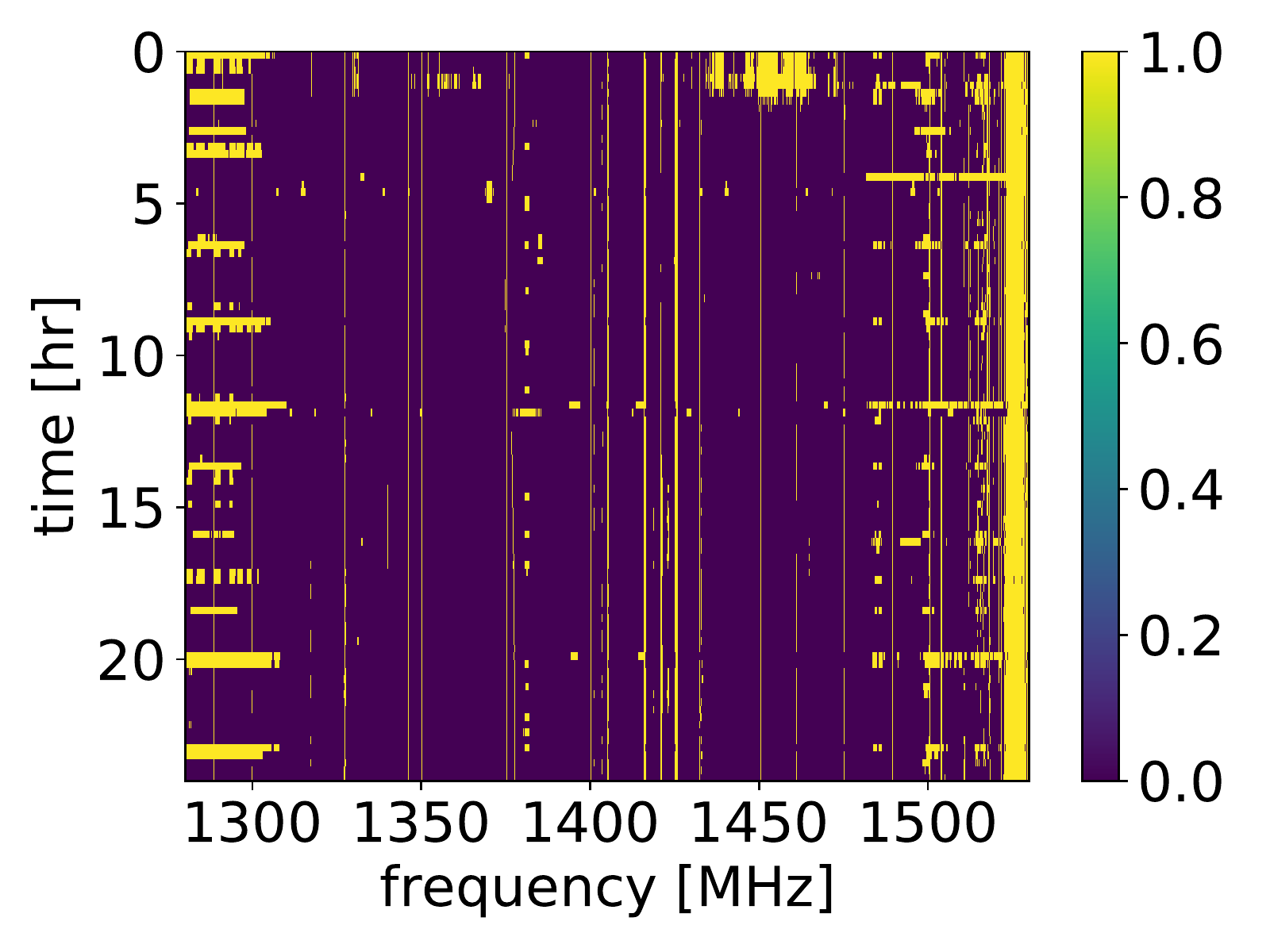}
		    \caption{A large fraction of the data are flagged and discarded for further astronomical processing.}
		    \label{fig:mitigated}       
        \end{subfigure}
\caption{RFI flagging and excision with data from the DSA-110 at Ownes Valley Radio Observatory (OVRO) \cite{hallinan2019dsa}     
} 
\label{fig:rfi}
\end{figure}

\begin{figure*}[h]
\centering
\begin{subfigure}[b]{0.3\linewidth}
    \includegraphics[width=\linewidth]{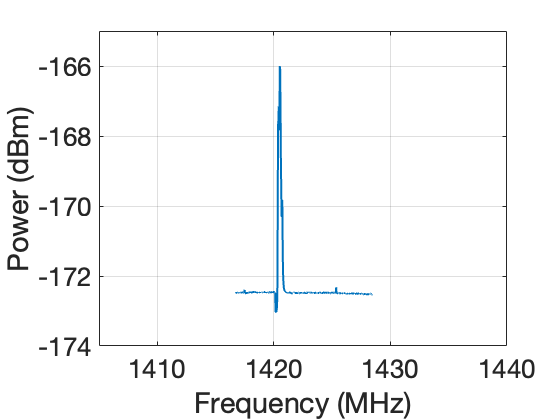}
    \caption{Galactic H1 line @ 1420 MHz}
    \label{fig:astro}
\end{subfigure}
\begin{subfigure}[b]{0.3\linewidth}
    \includegraphics[width=\linewidth]{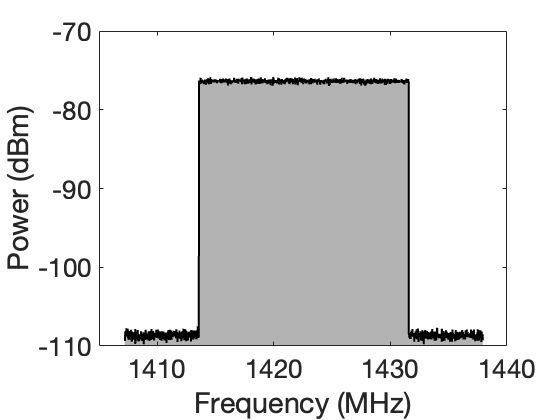}
    \caption{LTE signal at base station}
    \label{fig:lte_psd}
\end{subfigure}
\begin{subfigure}[b]{0.3\linewidth}
    \includegraphics[width=\linewidth]{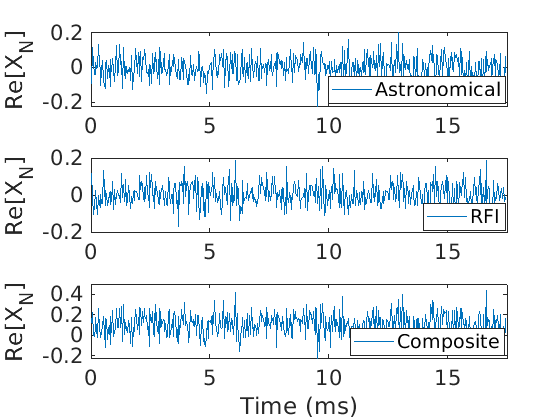}
    \caption{Signals in time-domain} 
    \label{fig:signals_time_domain}
\end{subfigure}
\begin{subfigure}[b]{0.3\linewidth}
    \includegraphics[width=\linewidth]{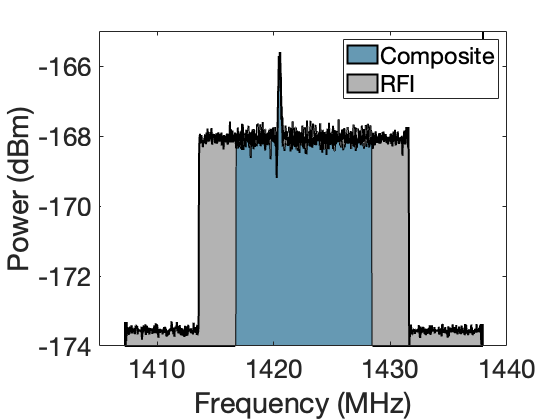}
    \caption{Composite signal with downlink RFI}
    \label{fig:composite_psd}
\end{subfigure}
\begin{subfigure}[b]{0.3\linewidth}
    \includegraphics[width=\linewidth]{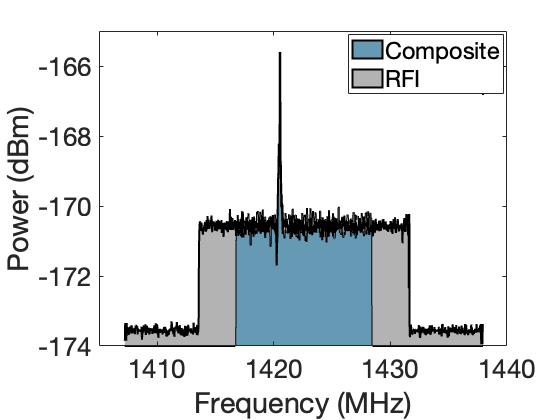}
    \caption{Composite signal with uplink RFI}
    \label{fig:composite_psd_ul}
\end{subfigure}
\caption{Spectral and temporal characteristics of Astronomical, RFI and Composite signals. RFI signal parameters for the downlink sample signal: 20 MHz bandwidth, 60\% frame occupancy,
RFI signal parameters for uplink sample signal: 20 MHz bandwidth, 60\% frame occupancy, contribution from 10 UEs.}
\label{fig:composite_sig}
\end{figure*}

\noindent {\textbf{Impact of RFI in Radio Astronomy:}} 
Probing the Universe at radio frequencies offers the possibility to study matter and energy under extreme conditions that cannot be achieved on Earth. For example, the omnipresent neutral atomic Hydrogen (HI) line, with rest frequency at 1420 MHz, has been extensively used to study the structure of galaxies and the intergalactic medium. As the universe is expanding, astronomical objects are moving away from us with increasingly high velocities, which creates a Doppler effect,
shifting all spectral lines from their rest frequencies to lower frequencies, a phenomenon termed ``redshift''. As a result, the HI line is now observed at frequencies around 1415 MHz from the Virgo cluster, at 1400 MHz from the Perseus supercluster, and at 1388 MHz from the Coma cluster of galaxies. The highly redshifted HI line (observed even at frequencies below 200 MHz) is also the unique probe available to study the distribution of matter in the early universe and to understand its epoch of reionization~\cite{Zaroubi_2012}.
Redshifted sources require astronomers to increasingly observe outside the protected bands~\cite{NAP21774, itu_radio_astronomy_13, craf_radio_astronomy_05}.
Moreover, astronomical emissions are extremely weak due to the large distances travelled before being detected by radio telescopes. Often, detection requires long integration at ${\approx}$ 40 dB below the telescope noise. This is in stark contrast to reception of a typical wireless communication signal at ${\approx}$ 20 dB above the receiver noise level. Hence, these receivers are extremely sensitive to RFI which may lead to false detection of the signal of interest or its masking, lowering the overall sensitivity of the telescope due to corrupted data excision. The RFI situation is worsening, mostly driven by two factors: population growth, increasing the chances of RFI polluting telescope sites, and the development of new commercial wireless technologies, like 5G and rural wireless broadband~\cite{fcc_rural,rau2019rfi,pawr-rural}. 
Figure \ref{fig:rfi} illustrates the excision problem with data collected with the Deep Synoptic Array DSA-110 \cite{kocz2019dsa,hallinan2019dsa} located at the Owens Valley Radio Observatory (OVRO), CA, USA (see \S \ref{subsec:acquisition}). Despite being geographically remote, it is affected by telemetry, communication and RADAR RFI. The data in Figure \ref{fig:contaminated} spanning 1280-1350 MHz is corrupted with three types of RFI encountered in radio astronomy:
continuous in time and narrow in frequency; intermittent in time and narrow in frequency; and impulsive in time and wide in frequency.
Figure \ref{fig:mitigated} shows the same data after identification and excision of the RFI-corrupted time-frequency bins. Here, the RFI detection consists of correcting the baseline of the averaged spectrum or time series using a median filter, then identifying the power excesses due to RFI on both time and frequency axis. The contaminated ranges of frequencies and times are then discarded. 
This flagging and excision approach is usually tuned to minimize the probability of non-detection of the RFI  
resulting in a significant data loss, sometimes as high as 40\% at L-band (1-2 GHz), 30\% at S-band (2-4 GHz) and 20\% at X-band (4-8 GHz) \cite{rau2019rfi}, impacting the sensitivity of the telescope and recovery of the astronomical signal of interest.


Therefore, even with state of the art methods in RFI mitigation, full recovery of an astronomical signal corrupted by RFI cannot be achieved without prior knowledge of the source of RFI. Fortunately, communication signals can be \textit{characterized} accurately and made available to the telescope through collaboration, which can be \textit{intelligently cancelled} from the telescope data to reveal the astronomical signal. Baseline results for such scenario with simple radio signals as contamination has been presented in our previous work~\cite{rfi_dyspan_2021}. In this paper, this hypothesis has been tested through the simulations and experiments presented  that includes injecting realistic LTE signals generated based on the 3GPP standards and realistic channel conditions for LTE signal propagation. It provides a comprehensive analysis of the proposed system along with extensive parameter space exploration covering majority of possible scenarios of LTE RFI contamination.

{To the best of our knowledge, no prior work applies a priori information of RFI to continuously cancel the interference at the telescope.}

%% file: prelim.tex
\section{Models and Preliminaries}
\label{sec:prelim}

\subsection{Acquisition and processing of astronomical signals} 
\label{subsec:acquisition}
A radio telescope achieves its high sensitivity by maximizing its directivity, collecting areas, and minimizing the system temperature of its receivers. It can vary from large single dish antennas equipped with single or multiple beam receivers, to large arrays of antennas that are either phased together to produce multiple beams in the sky or to perform interferometric synthesis imaging \cite{wilson2009tools,taylor1999synthesis}.
After signal conditioning (i.e., amplification, equalization and filtering), the output of the individual receivers are digitized over hundreds of MHz and channelized into smaller frequency 
bins of hundreds of kHz width~\cite{price2021spectrometers}. Channelization is useful 
for reducing the data rate for real-time processing, 
share computational resources 
and excise RFI-corrupted channels before further processing. 
Subsequent data processing are specific to the observed astronomical object and may include real-time matched-filtered transient searches, spectral integration or data correlation for synthesis imaging.

\noindent\textbf{The OVRO DSA-110 Radio Telescope:} 

\begin{figure}
\centering
\includegraphics[width=1\linewidth]{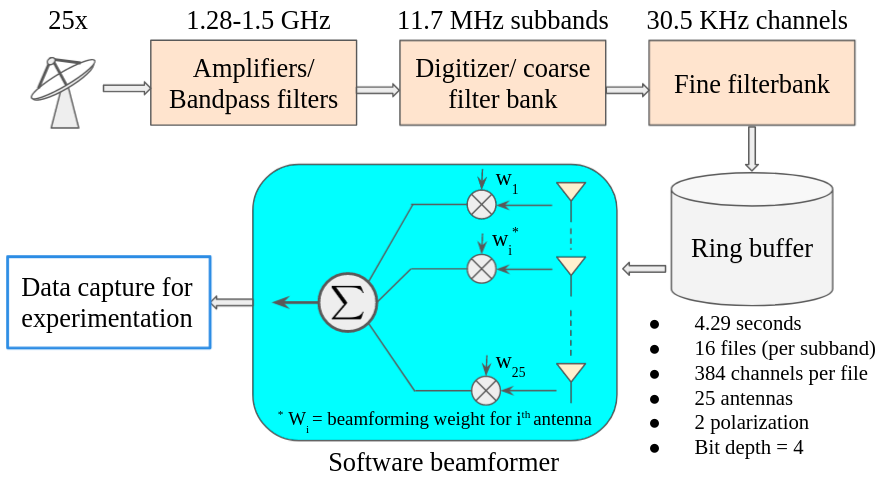}
\caption{Block diagram of signal capture and post-processing at telescope}
    \label{fig:astro_capture}
    \vspace{-15pt}
\end{figure}

The astronomical dataset, utilized throughout in this paper, has been collected with the Deep Synoptic Array being deployed at the OVRO. Figure~\ref{fig:astro_capture} depicts the processing steps involved in data collection and post processing at OVRO and are elaborated in \S~\ref{sec:experiment}. Figure \ref{fig:astro} shows the galactic H1 line at 1420 MHz, collected at OVRO, which is identified as the astronomical signal of interest in this work. It is important to note that the H1 line visible at 1420 MHz falls within a protected band dedicated to radio astronomy and does not contain real RFI. We have injected simulated LTE signals as RFI to the astronomical signal for experimentation purposes.

\subsection{Signal model for astronomical signals}

The digitized and channelized output of a single telescope antenna (either single dish or an element of an array) is expressed as:
\begin{align}
x_{T}[n] &= x_A[n] + x_N[n] + x_R[n]
\approx x_N[n] + x_R[n]
\label{eq:datamodel}
\end{align}
where $x_{T}[n]$ is the channelized baseband signal at a filterbank channel centered around frequency $f_c$ and at time sample $n$, and follows a stationary (assumed over a short duration) stochastic process, and is independently and identically distributed (i.i.d.) with $x_{T}[n] {\sim} \mathcal{NC}(x_R[n],\sigma^2)$. $\mathcal{NC}(\mu,\Gamma)$ indicates the circular complex Gaussian distribution with mean $\mu$ and covariance $\Gamma$. $x_A[n] {\sim} \mathcal{NC}(0,\sigma_A^2)$ is i.i.d. and represents the accumulated contribution of all astronomical sources in the field of view of the telescope, $x_N[n] {\sim} \mathcal{NC}(0,\sigma_N^2)$ is i.i.d. and represents the system noise contribution, and $x_R[n]$ is the deterministic RFI contribution with power $\sigma_R^2 {=} N^{-1}\sum_N|x_R[n]|^2$. However, accurate characterization of $x_{T}[n]$ and $x_R[n]$ is non-trivial and is discussed in \S\ref{sec:collab}. 

\subsection{Signal model for LTE RFI signal}
Long term evolution (LTE) signals employ a multicarrier modulation scheme to maximize spectral efficiency called 
orthogonal frequency division multiplexing (OFDM) with a variety of parameters defined by the 3rd generation partnership project (3GPP) standardization body. The general model for an OFDM signal for typical transmissions using a carrier at frequency $f_0$ is shown in \eqref{eq:ofdm}:
\begin{equation}
x_R(t)=\operatorname{Re}\left\{e^{{j} 2\pi f_{0} t} \sum_{k=-N_{\textrm{FFT}}/{2}}^{N_{\textrm{FFT}}/{2}} \alpha_{k} {e}^{{j} 2 \pi k\left(t-t_{{g}}\right) / T_{{u}}}\right\} 
\label{eq:ofdm}
\end{equation}

\setlength{\tabcolsep}{0.2em}
\begin{table}
\vspace{5pt}
\begin{center}
\scriptsize
\begin{tabular}{ |c|c|c|c|c|c|c| } 
\hline
\textbf{Bandwidth (MHz)}  & 1.25 & 2.5 & 5  & 10 & 15 & 20  \\
\hline
\textbf{Occupied BW (MHz)}    & 1.140 & 2.265 & 4.515  & 9.015 & 13.515 & 18.015  \\
\hline
\textbf{Frame (ms)}      & \multicolumn{6}{c|}{10}    \\
\hline
\textbf{Subframe (ms)}   & \multicolumn{6}{c|}{1}     \\
\hline
\textbf{Sampling Frequency (MHz)} & 1.92 & 3.84 & 7.68  & 15.36 & 23.04 & 30.72  \\
\hline
\textbf{$\boldsymbol{N_{FFT}}$}                 & 128 & 256 & 512  & 1024 & 1536 & 2048  \\
\hline
\textbf{$\boldsymbol{N_{subcarrier}}$}    & 76 & 151 & 301  & 601 & 901 & 1201  \\
{(Including DC subcarrier)}  &  & &   &  &  &   \\
\hline
\textbf{$\boldsymbol{N_{guard}}$ }    & 52 & 105 & 211  & 423 & 635 & 847 \\
\hline
\textbf{Resource Blocks}    & 6 & 12 & 25  & 50 & 75 & 100  \\
\hline
\textbf{OFDM Symbols/Frame}   & \multicolumn{6}{c|}{7/6 (short CP/long CP)}      \\
\hline
\end{tabular}
\end{center}
\caption{Different LTE parameters used for simulation}
\label{table:lte_spec}
\vspace{-15pt}
\end{table}

\begin{figure*}
\centering
\begin{subfigure}[b]{0.42\linewidth}
    \includegraphics[width=\linewidth]{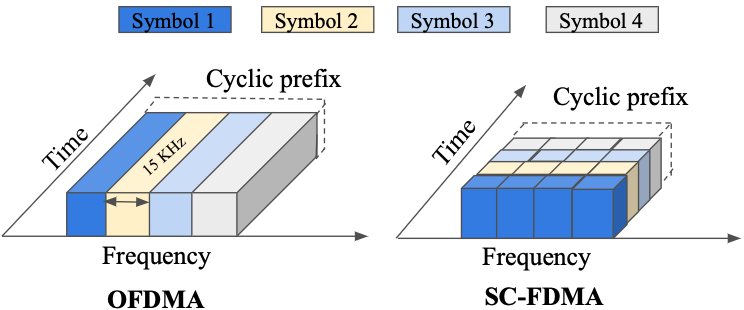}
    \caption{Time frequency grid comparison of downlink (OFDMA) ans uplink (SC-FDMA)}
    \label{fig:time_freq_grid_lte}
\end{subfigure}
\begin{subfigure}[b]{0.53\linewidth}
    \includegraphics[width=\linewidth]{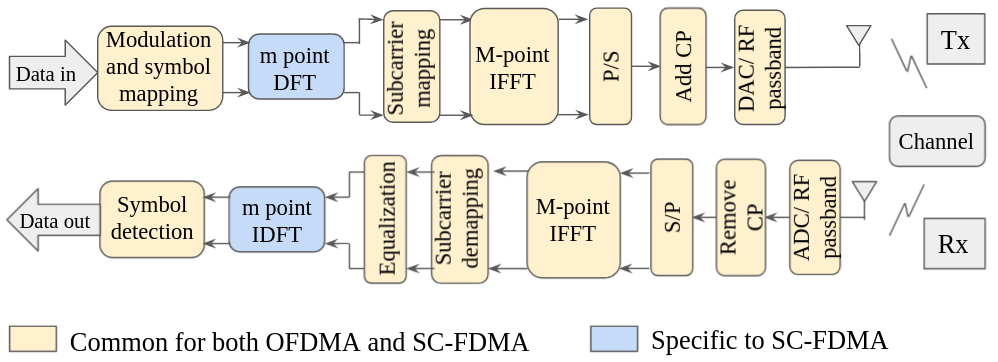}
    \caption{OFDMA and SC-FDMA Tx-Rx chain, blue DFT/IDFT block denotes additional encoding in SC-FDMA (m$<$M)}
    \label{fig:tx_rx_chain_lte}
\end{subfigure}
\caption{Uplink and Downlink LTE signal generation}
\label{fig:lte_gen}
\end{figure*}

Where $\alpha_k$ is the m-ary modulated symbol which is placed on $k_{th}$ subcarrier, $T_u$ is the symbol duration and $k$ is number of subcarriers. For downlink transmission, orthogonal frequency division multiple access (OFDMA) is employed, whereas, for uplink transmission purposes, single carrier frequency division multiple access (SC-FDMA) is used. Both of these methods are variations of the frequency division multiplexing scheme tuned based on the requirements and limitations in uplink and downlink. Figure~\ref{fig:time_freq_grid_lte} presents the subcarrier assignment scheme of modulated symbols for both OFDMA and SC-FDMA with a comparable reference frame including four symbols. In Figure~\ref{fig:tx_rx_chain_lte} we observe the difference in  transmitter receiver chain of both modulation scheme being - requirement of an additional DFT/IDFT stage in symbol generation/detection for SC-FDMA. It dictates the symbol assignment to subcarriers represented in Figure~\ref{fig:time_freq_grid_lte} mitigating the issue of high peak to average power ratio due to parallel transmission of different symbols as it can be detrimental for power limited User equipment (UE). In SC-FDMA, symbols are transmitted at a larger bandwidth and at a higher rate (smaller symbol duration in each subcarrier). This way the power variation remains low while maintaining the other benefits of OFDMA. 


The signal bandwidth depends on the size of the FFT, number of subcarriers, guard bands, etc. as described in Table \ref{table:lte_spec}.  These parameters are identical for OFDMA and SC-FDMA used in downlink and uplink. Typically, the LTE signal is grouped as radio frames in baseband with each frame containing 10 subframes and each subframe consisting of two slots. The resource block is the smallest unit of an LTE frame allocated to a user. Each resource block consists of 12 subcarriers lasting for a duration of 7 symbols. This arrangement along with the parameters in Table \ref{table:lte_spec} lend unique characteristics to the RFI. However, propagation over large distances and multipath reflections deteriorates the features of the RFI in time and frequency, hence the need for robust stochastic signal decomposition. An example realization of the downlink LTE signals used as RFI in this work is shown in figure \ref{fig:lte_psd}, which typically occupies a wider band than the telescope fine channel bandwidth. In reality, if the telescope acquires an astronomical signal at or around $f_0$, this RFI will undergo the same processing as mentioned in \S\ref{sec:prelim} and will be present in varying strength across multiple telescope channels depending on the telescope aperture, sidelobe gain and the spectral occupancy of the LTE signal itself. We refer to this signal as the \textit{composite signal}, which is defined by \eqref{eq:datamodel} and \eqref{eq:ofdm}. An example time-domain signal representation of the RFI and composite signal is depicted in Figure~\ref{fig:signals_time_domain}. The RFI appears stochastic in time due to the polyphase channelization of the telescope data. The power spectral density of the composite signal in presence of downlink RFI and uplink RFI from identical cell topology and distance from telescope but varying traffic patterns are presented in Figures \ref{fig:composite_psd} and \ref{fig:composite_psd_ul}. Average and maximum power levels and power control mechanisms are implemented for both uplink and downlink signal power in the spectral plots presented as per the 3GPP standards, but wireless channel effects are not displayed in those spectral plots. Additional information regarding the LTE signal generation for experimentation and validation purposes of the proposed method is provided in \S\ref{sec:experiment}.

%% file: system.tex
\section{Collaborative RFI cancellation}
\label{sec:collab}
The literature on cellular RFI mitigation through cancellation in radio astronomy primarily focuses on extracting signal features at the radio telescope via local sensing, without any prior knowledge of the source of the RFI (see \S\ref{sec:related} for literature review). As a result, the RFI signal information is extremely limited due to equipment constraints (front-end, BW, gain, etc.), signal deterioration due to propagation and multipath effects, and lack of coherence between the RFI source and the telescope. Another limitation stems from the very methods employed to characterize the RFI, which are almost always limited to Fourier methods for frequency domain analysis and temporal statistics like autocovariance, cyclostationarity, or higher order statistics. All of these methods are sensitive to time-varying RFI from cellular networks and require long observation times to accumulate a steady-state model. Furthermore, cancellation often requires local synthesis (requires sharing of user data) of the RFI signal from the acquired characteristics to either employ time-domain nulling (subtraction) or frequency domain filtering. This has the risk of eliminating the astronomical signal of interest and requires high degree of synchrony between the telescope and RFI sources for phase coherent cancellation. 

\begin{figure}
\centering
\includegraphics[width=\linewidth]{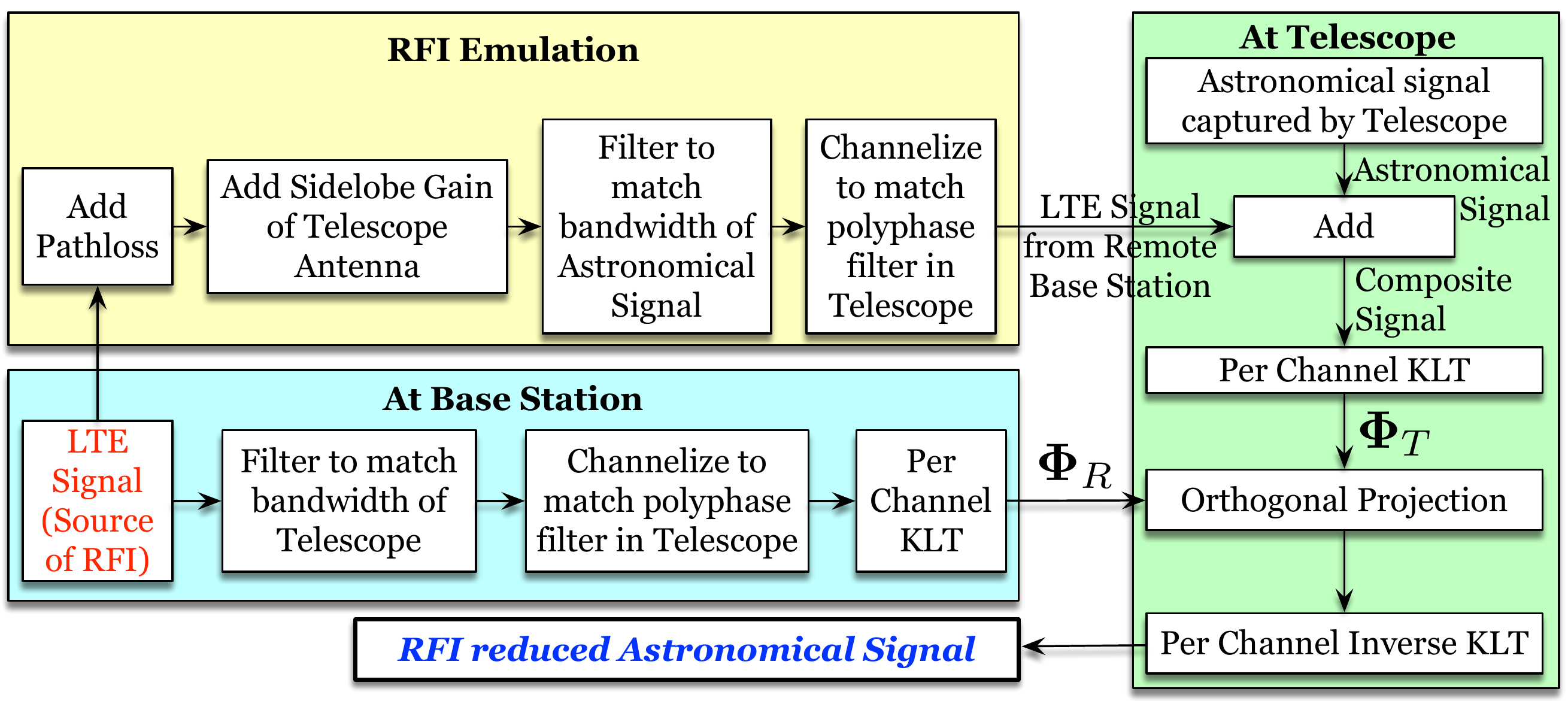}
\caption{Block diagram of the RFI cancellation system}
    \label{fig:block}
\end{figure}

Our method is radically different from the state of the art in three aspects: 1) The RFI is characterized at the cellular base station (BS) into a compact yet accurate eigenspace that is periodically shared with the telescope via a shared channel over the Internet. Unlike Fourier methods, which decomposes using complex exponential bases only, our method extracts the bases from the signal itself, which adapts with time-varying cellular RFI. Furthermore, the fidelity of the decomposition is vastly improved at high signal power, which is maximum at the BS. Figure \ref{fig:rqf_bs} in \S \ref{sec:result} discusses this unique feature; 2) At the telescope the composite signal is decomposed using the same method revealing its eigenspace that contain the RFI subspace, ideally orthogonal to the astronomical signal space; and 3) The shared RFI eigenspace is used to cancel the RFI from the composite eigenspace via orthogonal projections. Since the cancellation happens in the eigenspace, a final step to convert the eigenspace to the time-domain signal will reveal the RFI-free astronomical signal. Figure~\ref{fig:block} shows the components and flow of collaborative information between the telescope and the cellular network. For the purposes of this work, the RFI is emulated as an LTE signal and combined with the real astronomical signal as described in \S\ref{sec:prelim}\footnote{In this work, we consider 1 cellular BS and 1 radio telescope as the foundation for more complex topologies, to be addressed in future work.}. Our analysis and simulation is based on this composite signal but adheres to all the channelization and bandpass filtering as employed by the DSA-110 telescope. At the BS, the eigenspace for the LTE RFI signal is shown as $\mathbf{\Phi}_R$ in figure~\ref{fig:block} and that of the composite signal at the telescope is denoted by $\mathbf{\Phi}_T$, which are formally defined in \S\ref{sec:rfi}. These two are the key parameters required at the cancellation step along other topological information given in Table~\ref{tab:shared_param}.

\begin{table}[h]
\begin{center}
\scriptsize
\begin{tabular}{ |c|c| } 
\hline
\textbf{Static Parameters} & \textbf{Dynamic Parameters} \\
\hline
Observed frequency range $^\dagger$ & Eigenspace ($\mathbf{\Phi}_R$) \\
\hline
Polyphase filter subchannel $^\dagger$ & KLT window  ($L$)$^*$\\
\hline
\end{tabular}
\end{center}
\vspace{-5pt}
\footnotesize{$^\dagger$ parameters shared by telescope only. Others are cellular network only.\\$^*$ see~\S\ref{sec:rfi}.}
\caption{Shared parameters for collaborative cancellation}
\label{tab:shared_param}
\end{table}

Each of the parameters have specific roles in the cancellation apparatus and are explained in subsequent sections where applicable. 
The static parameters are constant and can be made available via a database lookup. The dynamic parameters change over time and require periodic sharing. The largest update in terms of size is Eigenspace information, $\mathbf{\Phi}_R$. For  the KLT window length = $L$ and number of time samples = $N$, each eigen function update for each subchannel of the polyphase filterbank will be $4 {\times} L {\times} L$ bytes of data. In our empirical evaluations, this amounts to 1 Mbytes of data per subchannel. However, it is possible to use the same $\mathbf{\Phi}_R$ for all the channels if the RFI bandwidth is larger than that of the telescope since the RFI is equally incident across the entire telescope bandwidth. Although, this may not be applicable for RFI sources with lower bandwidth than the telescope's field of view or if there are interference signals present with partial overlap across the telescope bandwidth causing frequency span mismatch between RFI and fine channels of the telescope. As we will observe later in \S~\ref{sec:experiment_results}, the frequency overlap within telescope bandwidth is quite likely for uplink signal present as RFI. It is also possible to limit the rate of sharing by replicating the averaging window of the telescope at the BS to smooth any temporal variation of $\mathbf{\Phi}_R$. Additionally these techniques require further theoretical and experimental evaluation that are beyond the scope of this paper, and are left as future work.

\subsection{Eigenspace representation of RFI and composite signals}
\label{sec:rfi}

\begin{figure}
\centering
\includegraphics[width=\linewidth]{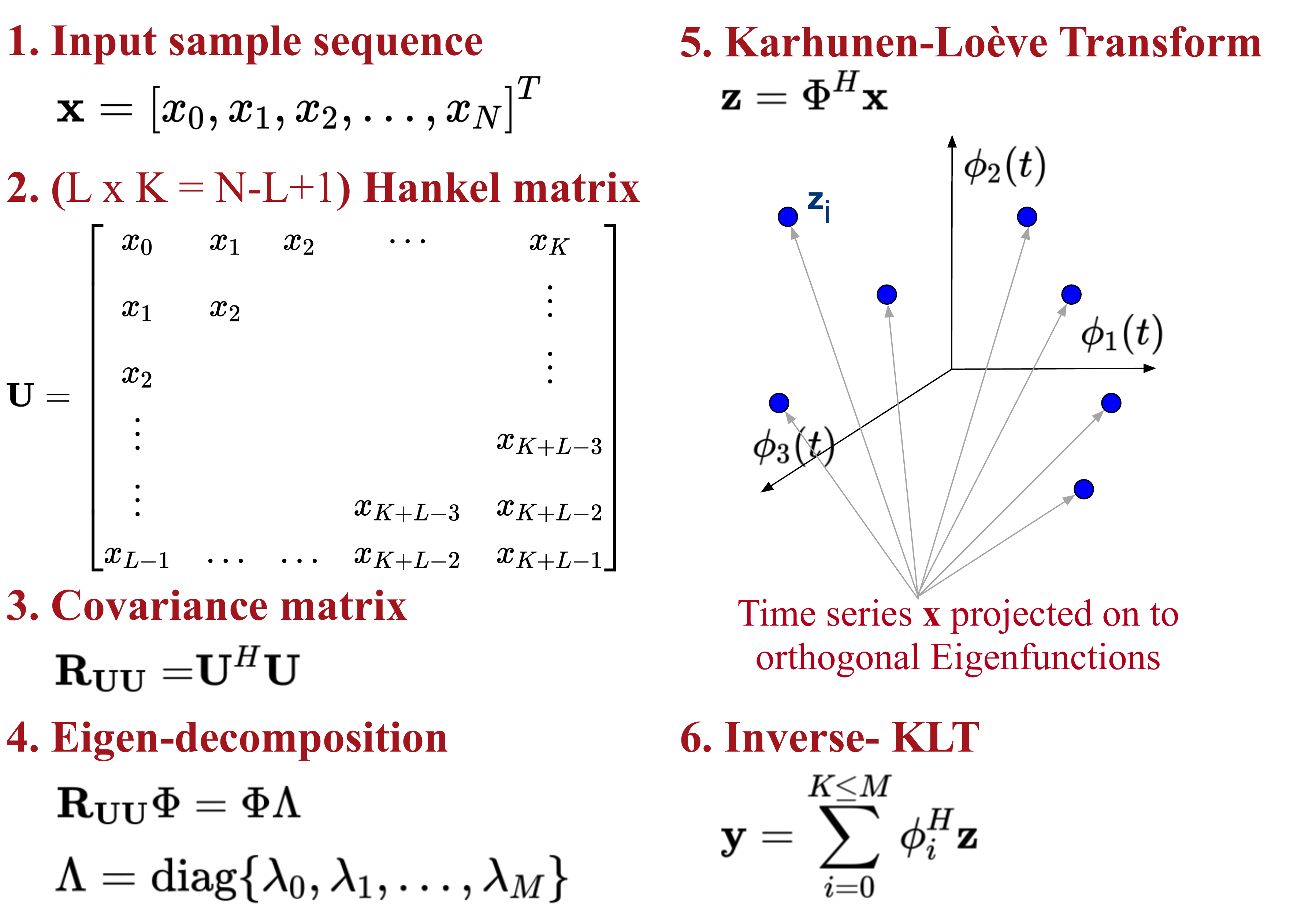}
\caption{Geometry of KLT based signal characterization
}
    \label{fig:klt}
\end{figure}
The Karhunen–Lo\`eve Transform (KLT) \cite{bem_book,RFI_KLT,trudu2020performance} decomposes any signal, and has the following advantages: (a) it is suitable for both narrow and wideband signals, (b) applicable to any type of basis and any type of signal (deterministic or stochastic), unlike other transforms like DFT that are limited to sinusoidal basis, (c) able to detect weak signals below the noise floor, and (d) optimal in the minimum mean square error sense. 
For analysis of sampled signal, KLT is often implemented using Singular Spectrum Analysis (SSA) that relies on the decomposition of the autocorrelation matrix, $\mathbf{R}_{xx}(n_{1},n_{2})$, which captures temporal correlations of a random signal $x[n]$, into its constituent eigenfunctions and corresponding eigenvalues. Let, $x[n]{=}x_T[n]$ at the telescope and $x[n]{=}x_R[n]$ at the BS, and $n_1$, $n_2$ are different time indices.
$\mathbf{R}_{xx}(n_{1},n_{2})$ is calculated as in \eqref{eq:Rxx} by embedding the signal $x[n]$ in a Hankel matrix, $\mathbf{U}$, of size ${L{\times} K}$~\cite{tome2018use,ssa} as shown in figure \ref{fig:klt}, where $N$ is the number of time samples in the signal of interest $x[n]$, $L$ is the KLT window size, empirically determined based on the astronomical signal being observed, telescope parameters and RFI characteristics and $L < K$ to prevent under-estimation of RFI subspace. 
\begin{align}
    \mathbf{R}_{xx}=\mathbb{E}[\mathbf{U}\mathbf{U}^H],\quad \text{where,} \quad \mathbf{U}=[\mathbf{x}_1,{\ldots},\mathbf{x}_K]
    \label{eq:Rxx}
\end{align}
where $\mathbf{x}_{i}{=}\left[x[n]{,}{\ldots}{,}x[n{+}L{-}1]\right]^{\mathrm {T} }$ are lagged vectors of size $L$, with $i{\in}[1,K]$, and $\mathbb{E}[.]$ is the expectation operator.
As a bottleneck of the collaborative apparatus, $L$ directly impacts the amount of data sharing and data rate required for success of the RFI cancellation apparatus. We have analysed the effects of varying KLT window lengths $L$ later in this paper to gain insight of the task of practical collaboration and provide a rudimentary budget of communication and memory resources required for that. Its analytical determination is out of scope of this work. 

Then, the KLT decomposition is obtained by solving the eigenvalue problem in \eqref{eq:eig_decomp}, 
\begin{align}
    \mathbf{R}_{xx}=\boldsymbol{\Phi} \Lambda \boldsymbol{\Phi}^H, \quad \text{where,} \quad \Lambda=\operatorname{diag}\left\{\lambda_{0}{,} {\ldots} {,} \lambda_{L-1}\right\}
    \label{eq:eig_decomp}
\end{align}
where $\lambda_{j}$ are the eigenvalues, with $j{\in}[0,L-1]$, and $\boldsymbol{\Phi}$ is a unitary matrix containing $L$ eigenvectors as its columns.
Since, \eqref{eq:eig_decomp} decomposes the temporal correlations of $x[n]$, each column of $\boldsymbol{\Phi}$, i.e. $\boldsymbol{\phi}_i$, is a time-series, and consequently is referred to as an eigenfunction.
Geometrically, the eigenfunctions give an orthogonal set of ``directions" (or spatial signatures) present in the autocorrelation matrix, which span the Hilbert space containing the KLT projected time samples, while the eigenvalues represent the power of the signal coming from the corresponding directions, sorted in decreasing order. 
Consequently, the KLT automatically adapts to the shape of the (signal$+$noise) irrespective of its behavior in time, by adopting a new reference frame spanned by the eigenfunctions, which makes it appropriate to characterize and subsequently remove RFI.
These eigenfunctions project the signal on to the Hilbert space given by, $\mathbf{z}_i{=}\boldsymbol{\Phi}^{H} \mathbf{x}_{i}$, where $\mathbf{z}_i$ are the columns of the matrix $\mathbf{z}$ and are orthogonal temporal principal components of the input $x[n]$ containing $L$ samples 
as shown in figure \ref{fig:klt}. 
Consequently, $\mathbf{z}$ is given by, $\mathbf{z}{=}\boldsymbol{\Phi}\mathbf{U}$ using the definition of $\mathbf{U}$ in \eqref{eq:Rxx}. 
Projecting $\mathbf{z}$ onto the $M({\le} L)$ eigenfunctions of $\boldsymbol{\Phi}$ with largest eigenvalues, reconstructs the signal, $\hat{{x}}[n]$ with minimum noise \cite{Szumski2011KLT} as shown in figure \ref{fig:klt}.
Depending on where the signal is characterized, at the telescope (includes added RFI and noise) or at the BS (RFI and noise only), the subscripts $T$ and $R$ are appended to the above variables, and the eigenfunctions are henceforth referred to as {\textit{Astronomical Kernel} ($\boldsymbol{\Phi}_T$}) and {\textit{RFI Kernel} ($\boldsymbol{\Phi}_R$)} respectively. 
$\boldsymbol{\Phi}_R$ provides an accurate and compact characterization of the RFI at the BS that is shared with the telescope for RFI cancellation as shown in figure \ref{fig:block}.
%

\begin{figure} 
\centering
\begin{subfigure}[b]{0.49\linewidth}
    \includegraphics[width=\linewidth]{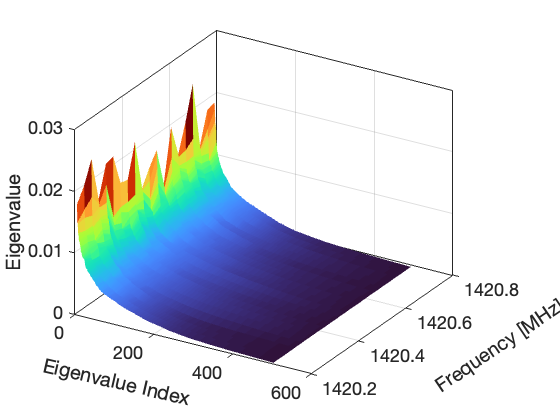}
    \caption{Composite signal at telescope}
    \label{fig:eigcomp}
\end{subfigure}
\begin{subfigure}[b]{0.49\linewidth}
    \includegraphics[width=\linewidth]{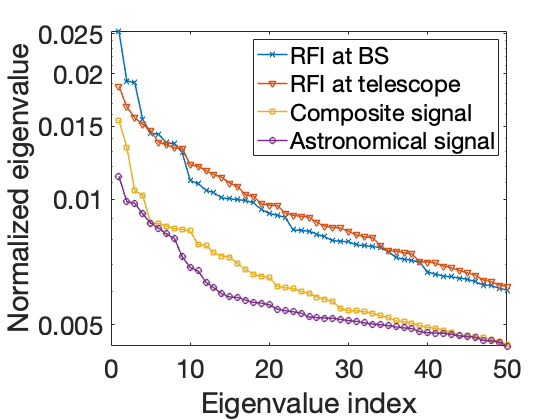}
    \caption{Eigenvalue comparison}
    \label{fig:eigval}
\end{subfigure}
\caption{Derived eigenspectrum from signal decomposition.\\
}
\label{fig:eig_spec}
\end{figure}

The eigenspectra of the decomposition of the composite signal at the telescope is shown in figure \ref{fig:eigcomp}. Figure \ref{fig:eigval} compares the 50 largest eigenvalues for the RFI at the BS, attenuated incident RFI at the telescope and the composite signal. The number of dominant eigenvalues in each plot shows the dimensionality of the associated signal subspace.

\subsubsection{Additional measures for uplink RFI decomposition} 
One of the crucial component of the proposed RFI cancellation apparatus is the characterization of RFI signal at its source. It is done so for downlink signals in baseband at the BS. Unfortunately that is not possible in case of uplink RFI for several reasons but not limited to: 1) UEs (source of uplink signal) do not have information about simultaneous uplink transmission, 2) UEs are power limited and lack the computation capability for decomposition and sharing the parameter sets of eigenfunctions, 3) efficient combining individual KLT outputs from UEs is hard to achieve. Thus, we propose the uplink signal decomposition to be performed at the respective BS the UEs are reporting to at a given time. Thus characterization is performed on the received signal at BS that is likely to include channel distortions and additive noise. During evaluation we have explored the consequences of such uplink RFI characterization.

\subsection{RFI cancellation with eigenspace projection}
\label{sec:cancel}

The KLT provides $L$ eigenfunctions at the telescope and the BS. Mathematically, RFI in the composite signal is cancelled by projecting the eigenfunctions at the telescope ($\boldsymbol{\Phi}_T$) onto a subspace that is \textit{orthogonal} to the subspace spanned by the RFI eigenfunctions ($\boldsymbol{\Phi}_R$).

\subsubsection{Orthogonal complement projector}
The first step in RFI cancellation is the computation of the orthogonal complement projector using the eigenfunctions estimated at the BS as in~\eqref{eq:orth_projection_matrix},
\begin{align}
\mathbf{P}_{\boldsymbol{\Phi}_R}^{\perp}{=}\mathbf{I}{-}\boldsymbol{\Phi}_R\left(\boldsymbol{\Phi}_R^{H} \boldsymbol{\Phi}_R\right)^{-1} \boldsymbol{\Phi}_R^{H}    
\label{eq:orth_projection_matrix}
\end{align}
where $\mathbf{I}$ is the $L{\times}L$ identity matrix. $\mathbf{P}_{\boldsymbol{\Phi}_R}^{\perp}$ is such that $\mathbf{P}_{\boldsymbol{\Phi}_R}^{\perp}\boldsymbol{\Phi}_R{=}0$ and 
by extension we can show that, $\mathbf{P}_{\boldsymbol{\Phi}_R}^{\perp}x_R[n]{=}0$ using the definitions in \eqref{eq:eig_decomp} and \eqref{eq:Rxx}. 
Therefore, applying this orthogonal projection
$\mathbf{P}_{\boldsymbol{\Phi}_R}^{\perp}$ to the telescope signal has the effect of nulling the RFI component.
The unique advantage of collaborative RFI mitigation is that the precision of the estimation of the RFI subspace, $\boldsymbol{\Phi}_R$, is improved by estimating it at the BS where the RFI is received at high SNR,
and consequently $\mathbf{P}_{\boldsymbol{\Phi}_R}^{\perp}$ can be calculated
even if the astronomical and RFI signals are not separable at the telescope.
Unlike the literature on signal separation or subspace estimation, which typically rely on the assumption of strong or weak signal separability or the existence of orthogonal subspaces of components in composite signals \cite{Golyandina2013SSABook}, this procedure does not require such assumptions, as the RFI subspace is accurately identified at high SNR at the BS.

\subsubsection{RFI cancellation}
The projection of the eigenfunctions at the telescope using the orthogonal complement projector in \eqref{eq:orth_projection_matrix} is given by \eqref{eq:rfi_cancellation}, 
\begin{align}
    \widehat{\boldsymbol{\Phi}}_T=\mathbf{P}_{\boldsymbol{\Phi}_R}^{\perp}\boldsymbol{\Phi}_T
    \label{eq:rfi_cancellation}
\end{align}
This projects the composite signal subspace at the telescope to the null-space of the RFI. Consequently, $\widehat{\boldsymbol{\Phi}}_T$ spans the subspace that is orthogonal to the RFI subspace spanned by $\boldsymbol{\Phi}_T$. This allows for subspace-based
removal of \textit{undesired} eigenfunctions corresponding to any RFI.

Finally, the inverse-KLT is used to reconstruct the RFI-free astronomical signal, i.e., $\hat{x}_T[n]$, which involves two steps.
First, the Hankel matrix corresponding to the RFI-free astronomical signal, $\widehat{\mathbf{U}}_T$, is reconstructed by projecting the matrix $\mathbf{z}_{T}$ onto the projected eigenfunctions $\widehat{\boldsymbol{\Phi}}_T$ 
as given by \eqref{eq:reconstruction}. 
\begin{equation}
\widehat{\mathbf{U}}_T=\widehat{\boldsymbol{\Phi}}_T \mathbf{z}_T, \quad \text{where,}\quad \mathbf{z}_T={\boldsymbol{\Phi}}_T^{H} \mathbf{U}_T
\label{eq:reconstruction}
\end{equation}
This cancellation method does not require the orthogonality of the astronomical and RFI subspaces in the composite signal at the telescope, since the projection in \eqref{eq:reconstruction} ensures that any RFI is nulled using the precise estimate of the RFI subspace at the BS.
Finally, the cross-diagonal elements of the reconstructed Hankel matrix are averaged \cite{Golyandina2013SSABook} using \eqref{eq:diagonal_averaging}, to reconstruct the space-signal time-series, $\hat{x}_T[n]$ from $\widehat{\mathbf{U}}_T$.
\begin{equation}
\hat{x}_T[n]= \begin{cases}
\frac{1}{n} \sum\limits_{k=1}^{n} \widehat{\mathbf{U}}_T^{(k{,}n{\text{-}}k{\text{+}}1)} &\text { for } 1 {\leqslant} n {<} L \\ 
\frac{1}{L} \sum\limits_{k=1}^{L} \widehat{\mathbf{U}}_T^{(k{,}n{\text{-}}k{\text{+}}1)} &\text { for } L {\leqslant} n {\leqslant} K \\ 
\frac{1}{N{\text{-}}n{\text{+}}1} \sum\limits_{\mathclap{k{\text{=}}n{\text{-}}K{\text{+}}1}}^{L} \widehat{\mathbf{U}}_T^{(k{,}n{\text{-}}k{\text{+}}1)} &\text { for } K{+}1 {\leqslant} n {\leqslant} N
\end{cases}
\label{eq:diagonal_averaging}
\end{equation}
where the superscript $(n{,}k{\text{-}}n{\text{+}}1)$ indicates the corresponding element in the matrix $\widehat{\mathbf{U}}_T$. 
This form of real-time recovery of $\hat{x}_T[n]$ is not possible with present methods in practice. Therefore, successful deployment of this method at telescope sites, like the DSA-110, will greatly reduce excision and maximize its sensitivity. In order to experimentally evaluate the quality of reconstruction of $\hat{x}_T[n]$, we define a metric that compares the residual  interference to the RFI-free astronomical signal.

%% file: rqf.tex
\section{RQF: A metric for evaluation}
\label{sec:rqf}

The performance of the proposed RFI mitigation approach is evaluated empirically using the Reconstruction Quality Factor (RQF) which measures the distortion in the recovered power after mitigating the interfering signal.

The RFI-mitigated signal at the telescope can be expressed as:
\begin{equation}
\hat{x}_T[n] = x_T[n] - \hat{x}_R[n]
\label{eq:rfifree}
\end{equation}
where $\hat{x}_R[n]{=}x_R[n] {+} \epsilon_r[n]$ is the estimated RFI contribution, and $\epsilon_r[n] {\sim} \mathcal{NC}(0,\sigma_{\text{est}}^2)$ captures the cumulative estimation and reconstruction error. The RQF over $N$ samples is defined as:

\begin{equation}
\text{RQF} = \frac{\| \tilde{x}_T - \hat{x}_T \|^2}{\|\tilde{x}_T\|^2}
\label{eq:rqf_new}
\end{equation}
where $\tilde{x}_T = x_A[n] + x_N[n]$ is the true RFI-free astronomical signal, and $\|x\|^2 = \frac{1}{N}\sum_N x^2[n]$ is the mean square error operator.

The advantage of defining the RQF this way is that it directly relates to the International Telecommunication Union (ITU) detrimental-level interference criterion \cite{raprotection}, which is set to 10\% power distortion of a 2000-seconds long integration. Assuming the independence of the RFI-free telescope time samples, this detrimental level can be adapted to the 4-seconds data sets produced by the telescope back-end. In that case, our proposed RFI mitigation approach reaches the ITU level when:
\begin{equation}
\text{RQF} \leq \frac{10 \%}{2000 \text{s}} \times 4\text{s} = 0.02 \%
\label{eq:ITUthreshold}
\end{equation}

%% file: results.tex
\section{Evaluation and Results}
\label{sec:experiment_results}

\subsection{Experimental Setup}
\label{sec:experiment}


\subsubsection{Astronomical Signal Acquisition}

\begin{figure}
\centering
\includegraphics[width=\linewidth]{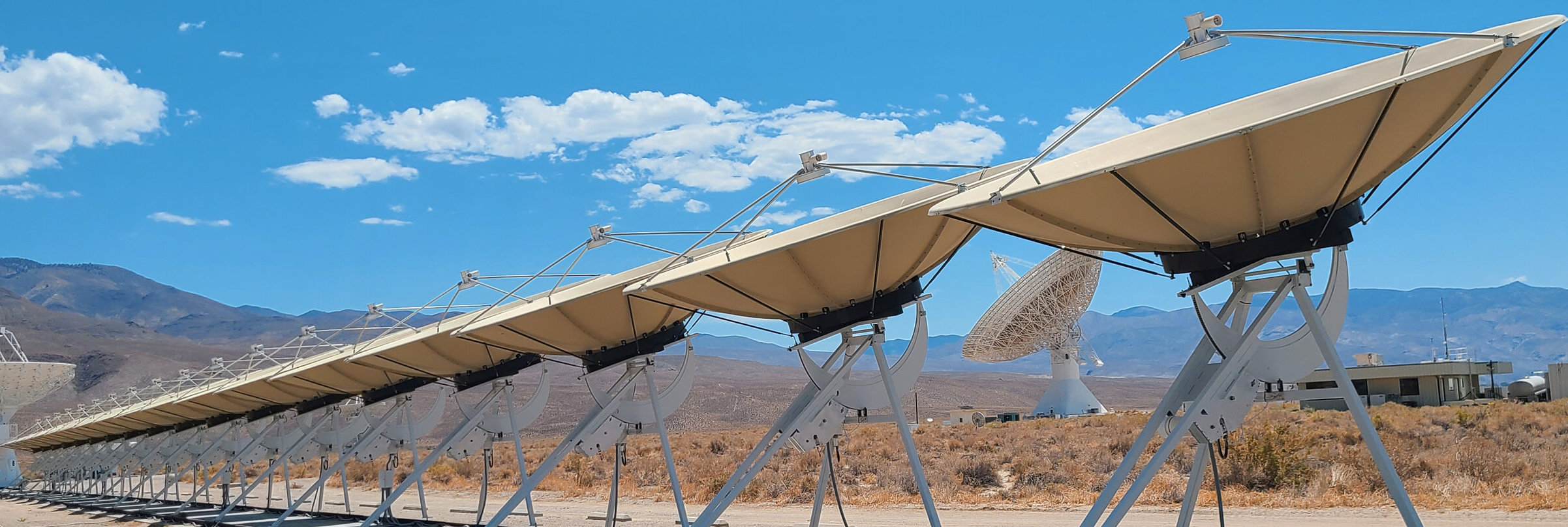}
\caption{The DSA-110 at OVRO.}
\label{fig:dsa110}
\end{figure}

The astronomical data used in this work have been collected on May 15 2021 with the Deep Synoptic Array (DSA-110, see Fig. \ref{fig:dsa110}), a radio interferometer made of the 110 4.65 m-antennas operating in the 1280 - 1530 MHz band, and located at the Owens Valley Radio Observatory near Bishop, CA~\cite{ovro}. At the time of data collection, only 25 antennas were built and operational. The DSA-110 is dedicated to the search of Fast Radio Bursts (FRB)~\cite{petroff2019fast}, and operates a real-time data processing pipeline to detect the bursts in beamformed data, record raw baseband data associates with them, and produce correlation matrices to localize their origin. The signal from each antenna is first amplified and filtered, then digitized and channelized into 11.7 MHz-wide coarse channels using a polyphase filterbank (PFB)~\cite{price2021spectrometers}.
The digitized coarse channels are then transferred over an Ethernet network to compute nodes in charge of forming beams within the field of view of the instrument, and searching these beams for FRBs using an incoherent de-dispersion search algorithm~\cite{walsh2018searching}.  This pipeline is implemented using a ring buffer architecture ~\cite{van2021psrdada}, and the baseband data for all antennas and coarse channels for a duration of 2 seconds contained in the buffers is written to disk in case of detection for further processing.

We manually triggered the download of baseband data and formed an arbitrary beam to simulate the output of a highly-directional single dish radio telescope for this work. We particularly focused on the coarse channel centered at 1420 MHz containing the galactic HI spectral line as a signal of interest \cite{dickey1990hi}.

\subsubsection{LTE Frame generation}
\label{sec:lte_transmit_gen}
\greg{}

\begin{figure*}[h]
\centering
\includegraphics[width=0.8\linewidth]{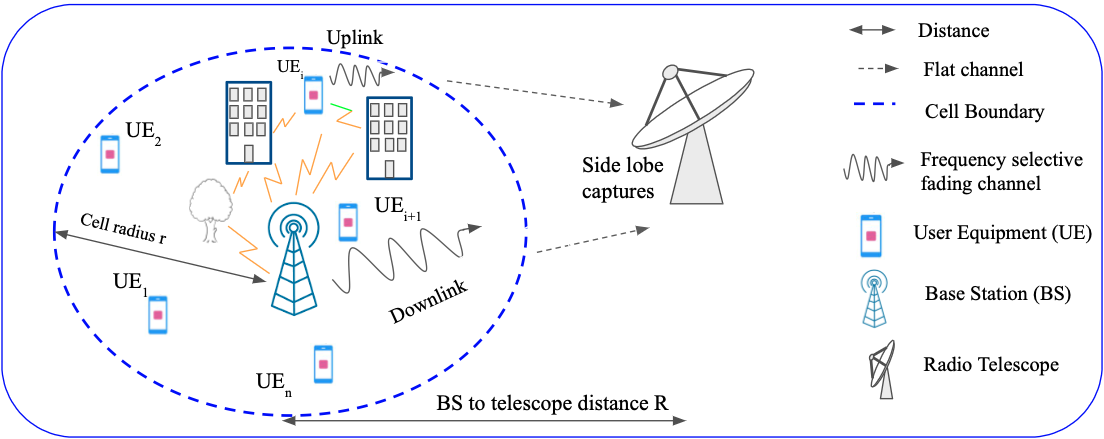}
\caption{LTE downlink and uplink RFI from a cell}
\label{fig:lte_block}
\end{figure*}

The LTE RFI signal is generated for both uplink and downlink transmission according to the parameters in Table \ref{table:lte_spec} and the specific transmission schemes laid out in \S\ref{sec:prelim}, containing 400 frames (10 milliseconds each) to be comparable in duration to the astronomical dataset. Three different modulations: QPSK, QAM-16 and QAM-64 are used in the LTE signals. The LTE frames contain primary (PSS) and secondary (SSS) frame synchronization signals. Other additional components (e.g, control channels) for LTE signal are not addressed for simplicity. Figure~\ref{fig:lte_block} presents structure of an LTE cell with uplink and downlink transmission along with the transmission links observed between the distant radio telescope and the BS and the UEs. We assume the cell radius is r kilometers whereas distance from BS to the telescope is R ($>$ r) kilometers. The Figure also depicts the side lobe capture of RFI at geographically isolated telescope location. For both downlink and uplink, UEs are uniformly distributed acrross the LTE cell with identical topology, but with different traffic pattern and we have introduced partial overlap of multiple uplink transmission band with the telescope bandwidth. 

\subsubsection{LTE transmission}
\label{sec:lte_transmit_exp}
 Transmit powers for downlink and uplink are maintained at different level based on the 3GPP standards. Downlink transmit power is fixed at 46 dBm, equally spread across the transmission band. In practice, Uplink transmission is power controlled with complex open-loop and closed-loop power control protocols in place due to the power limited UEs. We have implemented a simpler version of the power control protocol with maximum UE transmit power ($P_{max}^{UE}$) =  23 dBm at cell boundary with transmit power ($P^{UE}$) reducing towards the cell center inversely proportional to the square of distance from the BS (aligned with the propagation loss) for validation purposes of the proposed system. A mathematical representation of the power control protocol is presented in the following equation: $P_n^{UE} = P_{max}^{UE}.(\frac{d_n}{d_{max}})^2$, where $d_n, d_{max}$ are the distances of the $n_{th}$ UE from the BS and the cell radius respectively. We have validated our proposed RFI cancellation apparatus primarily with downlink signal, as evidences from telescope data suggests they are more likely to contaminate radio telescope data due to higher transmission power and directed beam patterns. The primary simulation is done with additive white Gaussian noise (AWGN) channel in place along with propagation loss between the RFI source and radio telescope. In later steps we performed more simulations with the following terrestrial channel models for both uplink and downlink to verify practical feasibility of the proposed apparatus: 1) urban micro cell, 2) bad urban macro cell, 3) indoor to outdoor channel, 4) urban macro cell, 5) bad urban macro cell, 6) suburban macro cell, 7) rural macro cell and 8) flat Rayleigh channel. Channel models 1 to 7 are from the WINNER II channel models~\cite{winner2_channel} along with their nomenclature. The RFI signals experience specific channel conditions within the cell and mostly flat channel beyond the cell boundary due to the geographical isolation of radio telescopes. In uplink scenario, channel experienced by an UE is relative to its location in the cell. For example, UE$_i$ in figure~\ref{fig:lte_block}, experiences the specified channel for a much smaller distance due to its proximity to cell boundary and experiences flat channel for rest of propagation path to the telescope. The corresponding uplink signal used to generate characterization parameter ${\boldsymbol{\Phi}}_R$ traverses a longer path through the specified channel up to the BS. On the other hand, signal from UE$_{i+1}$ travels a longer path through the channel specified up to cell boundary while propagating to the telescope, but a much shorter uplink to BS. Such variation in channel deterioration and propagation loss amounts to a diverse trend in RQF, unlike the downlink RFI scenario where signals are transmitted from a fixed location of the BS and are characterized in baseband.

\subsubsection{RFI injection at Telescope}
\label{sec:rfi_inject}
The LTE signal is filtered post introduction of channel and propagation loss, to match the bandwidth of the telescope and attenuated by free space pathloss and side-lobe gains (RFI is primarily acquired via telescope side-lobes). The side lobe attenuation depends on the angle of elevation of the antenna. The value of telescope gain can thus be anywhere between $10.log10(Ae*(pi*D\lambda)^2) =$ 15.1 dBi to $-\infty$ dBi, depending on whether the RFI is received through the main lobe or any null in the directivity of the dish (Ae = 0.7, D = 4.65, $\lambda$ = 21 cm). An arbitrary value of side-lobe gain of the telescope is chosen as $= -40$ dBi at OVRO DSA-110 for the simulation purposes. 
The BS location for the simulation is determined to be at $20$ km ($\sim$ 92.46 dB propagation loss) linear distance from the telescope, which is very close to the estimated distance of the closest BS to DSA-110. This brings the LTE signal to a power level such that it is comparable to the astronomical signal as shown in Figure~\ref{fig:composite_psd}. The composite signal spectrum from uplink RFI in Figure~\ref{fig:composite_psd_ul}, depicts overlap of RFI with different center frequencies, thus covering the guard bands of each other. The spectral plot is limited to the same bandwidth as the downlink rfi spectral plot to provide visual comparability. The signal is channelized (subchannels of telescope back-end post processing, not to be confused with the wireless channel) both at the BS and the telescope to match the channels of the telescope, which is a static parameter shared by the telescope as in Table \ref{tab:shared_param}. 
The attenuated and noisy signal is added to the astronomical signal to simulate the RFI contaminated composite signal. 

\begin{figure}
\centering
\includegraphics[width=0.75\linewidth]{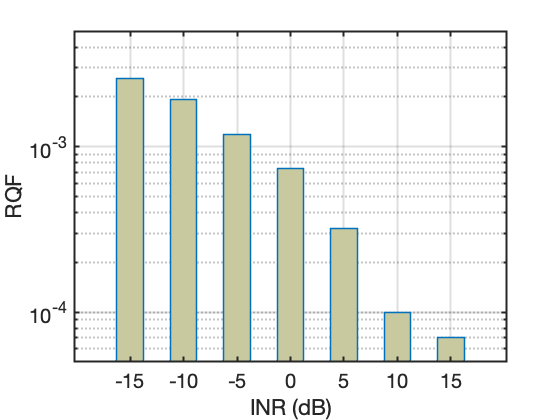}
\caption{Reconstruction loss of RFI with varying INR.} 
    \label{fig:rqf_bs}
\end{figure}

\subsubsection{Signal decomposition and reconstruction}
The LTE signal at BS (both uplink and downlink as discussed before) and the composite signal at the telescope are decomposed to obtain the eigenfunctions based on the formulation in \S\ref{sec:collab}. In our SSA based implementation, we limit the KLT window length to a finite value for the Hankel matrix and approximate the eigenspace to contain eigenfunctions corresponding to eigenvalues within 1\% of the maximum eigenvalue. This value of L is expected to determine the longest periodicity captured by SSA, which should be longer than the signalling rate of interference signal. Here it is empirically estimated to be L = 500 (16.412 $\mu$s). RFI characterization and reconstruction with each set of wireless channel scenario mentioned in\S\ref{sec:lte_transmit_exp} is performed 100 times and the reconstruction quality is averaged over the 100 simulations to smooth out any outlier behavior of the proposed apparatus.
Figure \ref{fig:rqf_bs} shows the reconstruction accuracy of a sample LTE signal with finite eigenspace using SSA \cite{rqf_ssa}
as a function of the interference to noise ratio (INR). We observe that the reconstruction accuracy improves with INR, meeting the RQF criterion presented in \S\ref{sec:rqf} at 10 dB INR. This is the main motivation for characterizing RFI at the BS instead of at the telescope, as the telescope receives the RFI at a much lower INR due to propagation loss leading to improper eigenspace representation and erroneous cancellation.
 Time synchronization between the source of RFI and the telescope is also of utmost importance for such collaborative apparatus. From Figure~\ref{fig:lte_block}, arrival time of a downlink RFI at the telescope is (t$_{\textrm{p}}$+R/c), t$_{\textrm{p}}$ is the processing time for characterization of the RFI signal. With R = 20 km, the arrival time = (66 $\mu$s + t$_{\textrm{p}}$) $\mu$s. The same for an UE at a distance r from the BS becomes ((R+r)/c + t$_{\textrm{p}}$). In this work, we assume that the collaboration between BS and telescope will start using standardized GPS clocks at both locations and appropritate delays will be introduced at the telescope accordingly.
Any effects on the proposed RFI mitigation metric owing to further synchronization error is explored later in terms of changes in reconstruction quality.

\setlength{\tabcolsep}{0.2em}
\begin{table}
\begin{center}
\scriptsize
\begin{tabular}{ |c|c|c| } 
\hline
\textbf{Properties}  & \textbf{Uplink} & \textbf{Downlink}   \\
\hline\hline
\textbf{Channel Bandwidth (MHz)}     & \multicolumn{2}{c|}{\{10, 15 , 20\}}  \\
\hline
\textbf{Distance from BS to the telescope (km)}      & \multicolumn{2}{c|}{20}    \\
\hline
\textbf{Power control}   & Yes & No     \\
\hline
\textbf{Frame occupancy (\%)}                  & {30 - 80}   & {30 - 80}\\
\hline
\textbf{No. of UEs }    & 5 (within a given cell radius) & --  \\
\hline
\end{tabular}
\end{center}
\caption{Different LTE parameters produce unique RFI}
\label{table:lte_sim_spec}
\end{table}


\subsection{Experimental Results}
In this section we have reported our findings using downlink signal as RFI propagated through AWGN channel that enables us to establish a baseline for the measure of accuracy of the proposed RFI cancellation method. We have additionally performed parameter space exploration with this downlink RFI signal to investigate the effects of factors like interference to noise ratio (INR) at telescope, frame occupancy in LTE RFI and time synchronization error between RFI source and telescope. The results for these are presented in Figure~\ref{fig:rqf} in purple. The constant parameters (2 out of 3 in each scenario) are set as follows - 1) INR (ratio of RFI and
and noise power): 5 dB, 2) frame occupancy: 70\%, 3) Synchronization error: 0 second. The scale is marked as $\boldsymbol{\Phi}_R$ at BS, which indicates $\boldsymbol{\Phi}_R$ - the RFI eigenspace is obtained after characterization at BS.
Later, we have produced graphs presenting variation in reconstruction quality pertaining to the LTE channel and power control effects along with the impact of choice of the KLT window-length $L$ for producing the Hankel matrix.

\label{sec:result}
\subsubsection{Reconstruction of RFI-free astronomical signal} we obtain, the rectified astronomical signal as shown in Figure~\ref{fig:reconstruct_sig} along with the true astronomical signal and the composite signal, following the signal reconstruction steps described above. The power levels are relative to the measured noise floor at the telescope (-174 dBm) as the baseline (-8 dB marker). We achieved an RQF of 4.132$\times$10$^{\textrm{-4}}$ for this reconstructed astronomical signal at an estimated interference to noise ratio (INR) of 5 dB at the telescope.

\subsubsection{Effect of Interference to Noise Ratio (INR)} 
\begin{figure}
\centering
\includegraphics[width=0.75\linewidth]{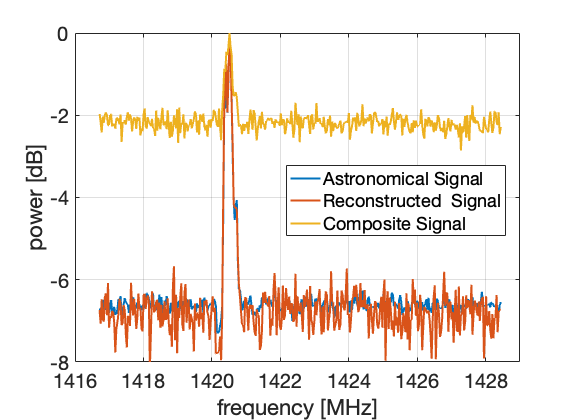}
\caption{Reconstructed space signal compared to the true astronomical and the composite signal.
}
    \label{fig:reconstruct_sig}
\end{figure}

A low INR is expected at telescope since RFI is generally acquired by the telescope side-lobes. For the reconstructed astronomical signal shown in Figure~\ref{fig:reconstruct_sig}, the INR is set at 5 dB as reported previously. In most practical cases, this will change based on the distance between telescope and the source BS and the varying side lobe attenuation due to change in elevation angle. We do not mention changes in  angle of azimuth as the radio telescope under consideration has only one degree of freedom (elevation) and fixed angle of azimuth. We have investigated the effect of INR varying from -10 dB to 20 dB on reconstruction quality using the RQF metric, shown in Figure~\ref{fig:rqf_inr}. RQF decreases with increasing INR as higher RFI power results in better characterization of the RFI facilitating better reconstruction quality~\cite{boonstra2005RFIthesis}. RQF reaches the ITU mandated threshold at 20 dB INR.

\subsubsection{Effect of synchronization error} 

GPS clocks have maximum time-synchronization error of 30 nanoseconds as reported in~\cite{GPSgovGP35:online}. Neglecting the processing time at the BS and transmission time of the RFI characterisation information compensated through the introduction of appropriate delays at the telescope, at most 4 samples of duration 32.826 ns can be out of synchronization. Effect of time synchronization error between BS and telescope of up to 320 nanoseconds is shown in Figure~\ref{fig:rqf_outOfSync}. Due to this synchronization error, RQF goes up to 4.6$\times$10$^{\textrm{-4}}$.

\subsubsection{RQF for varying spectral occupancy of the RFI} 
\begin{figure*}
\centering
  \subcaptionbox{Varying INR \label{fig:rqf_inr}}[.32\textwidth][c]{%
    \includegraphics[width=.33\textwidth]{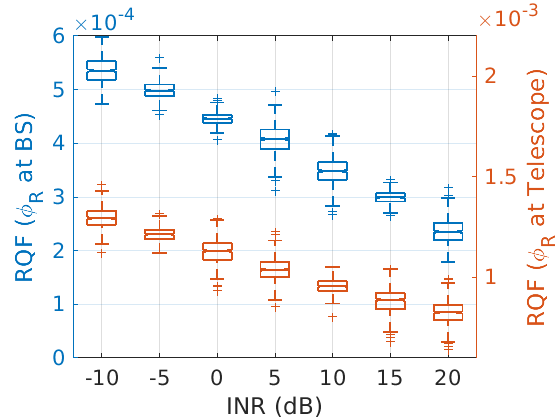}}
  \subcaptionbox{Varying $\mathbf{\Phi}_T$ lag time in samples
  \label{fig:rqf_outOfSync}}[.33\textwidth][c]{%
    \includegraphics[width=.33\textwidth]{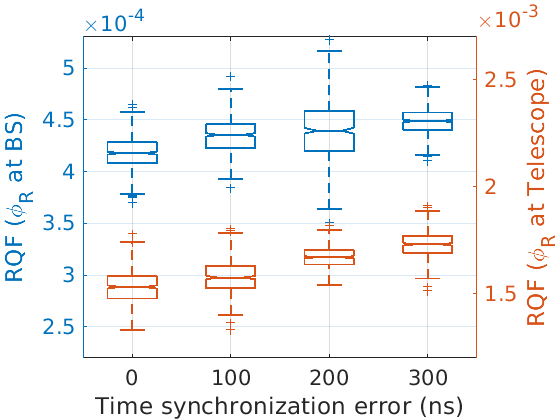}}
  \subcaptionbox{Varying RFI spectral occupancy \label{fig:rqf_occupancy}}[.33\textwidth][c]{%
    \includegraphics[width=.33\textwidth]{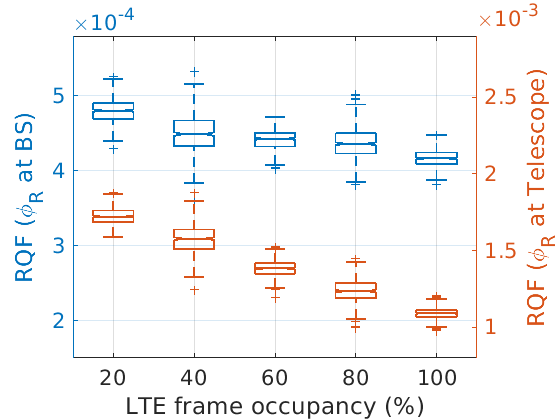}}
\caption{Reconstruction Quality Factor (RQF) variation caused by different RFI parameters}
\label{fig:rqf}
\end{figure*}

In a practical situation, LTE frames are not fully occupied and many resource blocks in a frame remain empty. The reconstructed signal in Figure~\ref{fig:reconstruct_sig} had an RFI consisting of 70\% occupied LTE frame. We have observed that with varying spectral occupancy, reconstruction quality of the astronomical signal changes as well. With different spectral occupancy, the eigenspace of the RFI signal gets skewed which produces higher error while projecting the composite signal and increases RQF indicating a poor reconstruction quality. Figure~\ref{fig:rqf_occupancy} shows the trend in RQF with changes in spectral occupancy of RFI. RQF improves with higher spectral occupancy but due to a low INR it is restricted to 4.1$\times$10$^{\textrm{-4}}$.

\subsubsection{RQF if RFI is decomposed at the telescope} 

Characterizing RFI at its source (BS for our use case), i.e at higher power level, is the most crucial and unique component of the proposed method which leads to improved RFI mitigation and astronomical signal reconstruction. To confirm this claim, we attenuated the emulated LTE signal to power levels comparable to the RFI at the telescope and followed the steps from \S~\ref{sec:experiment} to analyse all three scenarios mentioned above. 
These are expected to provide results equivalent to RFI signals captured locally at telescope site using a reference antenna for characterization and cancellation purposes. This method of locally characterizing RFI at telescope has been incorporated in several state of the art RFI mitigation techniques. We can observe the red plots in Figures~\ref{fig:rqf_inr}, \ref{fig:rqf_outOfSync} and~\ref{fig:rqf_occupancy} with the scales and axes marked as $\boldsymbol{\Phi}_R$ at Telescope indicating $\boldsymbol{\Phi}_R$ - the RFI eigenspace is obtained after characterization at telescope. The RQF is going up to 3$\times$10$^{\textrm{-3}}$ when RFI is characterized at telescope. Such error margins are higher in orders of magnitude higher than the expected threshold calculated in \S\ref{sec:rqf}, rendering the cancellation apparatus questionable. This demonstrates that, RFI characterization at the source of RFI leads to improved RFI cancellation and signal reconstruction.

\subsubsection{Effect of LTE practical channels}
\label{sec:lte_practical_chan}
In this section variations of RQF in different practical channel conditions are presented for both downlink and uplink RFI. Additionally, effects of parameter variation of the RFI cancellation and reconstruction apparatus on them are analysed.

\begin{figure*}[h]
\centering
\begin{subfigure}[b]{0.32\linewidth}
    \includegraphics[width=\linewidth]{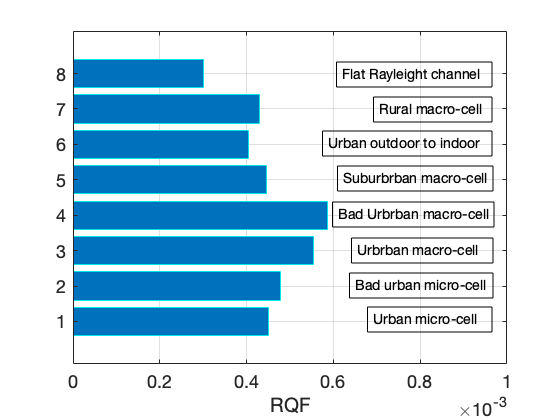}
    \caption{Downlink Scenarios}
    \label{fig:dl_rqf}
\end{subfigure}
\begin{subfigure}[b]{0.32\linewidth}
    \includegraphics[width=\linewidth]{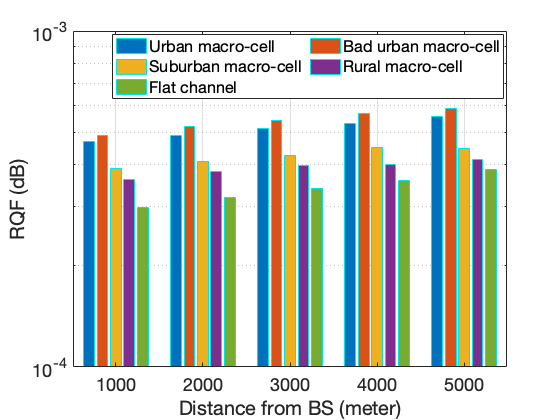}
    \caption{Uplink Macro-cell Scenarios}
    \label{fig:ul_rqf_macro}
\end{subfigure}
\begin{subfigure}[b]{0.32\linewidth}
    \includegraphics[width=\linewidth]{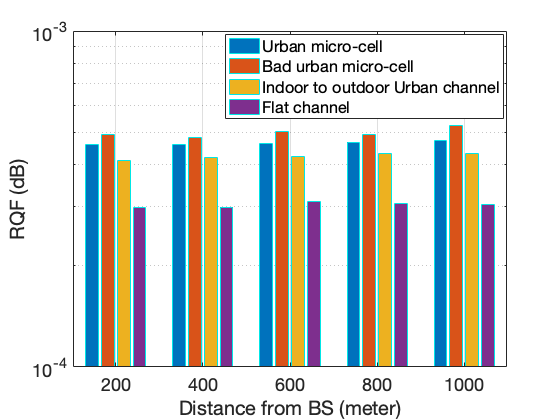}
    \caption{Uplink Micro-cell Scenarios}
    \label{fig:ul_rqf_micro}
\end{subfigure}
\caption{Reconstruction Quality in presence of RFI from Downlink and Uplink Signal - Different Channel scenarios}
\label{fig:uplink_rqf}
\end{figure*}

\paragraph{\textit{Downlink}}
Figure~\ref{fig:dl_rqf} shows the reconstruction quality for all eight different channel scenarios mentioned in \S\ref{sec:rfi_inject} with the previously described topology of the LTE cell, UEs and the radio-telescope in Figure~\ref{fig:lte_block}. We observe, micro-cells with a smaller cell radius and macro-cells with less severe channels like suburban or rural scenarios have RQF closer to the ITU mandated threshold, whereas RQF goes up to 6$\times$10$^{-\textrm{4}}$ for their counterparts with more severe channel and larger cell radius. In case of rural macro-cell, though channel effect and non-linearities are less severe, those are predominant in majority of the propagation path due to its much bigger cell radius. This leads to comparatively poor reconstruction quality with respect to other scenarios. 

\paragraph{\textit{Uplink}}
Effect of practical channels in case of uplink RFI and corresponding RQFs are shown in Figures~\ref{fig:ul_rqf_macro} and \ref{fig:ul_rqf_micro}. Macro-cells and micro-cells have different radius based on the channel conditions they are defined on. For example, a macro cell radius is $\sim$5000 meters and a micro-cell radius is $\sim$1000 meters. To provide comparable results in terms of distance of the UE from BS, we grouped the RQF values accordingly and presented the RQF variation based on distance from BS up to the cell boundary. This is relevant, because of the effect of propagation loss of uplink signal reaching the BS, that dictates the reconstruction quality. Higher propagation loss leading to lower INR reduces the decomposition and the reconstruction quality as we have already shown. On the other hand, the uplink signal captured at BS, contains the information about statistical properties of the channel, which is eventually transferred to the eigenspace representation of uplink signal. This information aids in the cancellation process as similar channel is observed by the RFI transmitted to the telescope up to the cell boundary. This effect can be observed in the rural microcell scenario, where unlike downlink, the RQF is better than most other scenarios. This is expected, because the propagation path mostly within the cell modifies the RFI with channel effects which are already present in the RFI subspace characterized at BS.
Additionally we observe less variation in RQF with increasing distance from BS in micro-cell, compared to that of macro-cell due to the difference in range of distance from UE to BS at which it is varied for micro-cell and macro-cell. But, range of values of RQF at the distance of 1000 meters in both micro-cell and macro-cell scenarios are consistent.

\subsubsection{Impact of KLT window length}
\label{sec:impact_intrinsic_param}
\begin{figure*}[h]
\centering
\begin{subfigure}[b]{0.32\linewidth}
    \includegraphics[width=\linewidth]{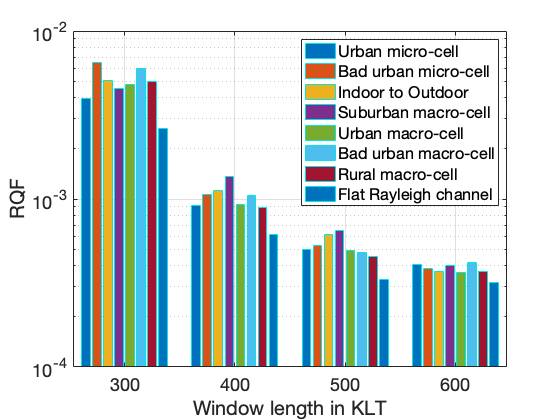}
    \caption{Reconstruction quality estimate over KLT window length - Uplink}
    \label{fig:rqf_ul_winlen}
\end{subfigure}
\begin{subfigure}[b]{0.32\linewidth}
    \includegraphics[width=\linewidth]{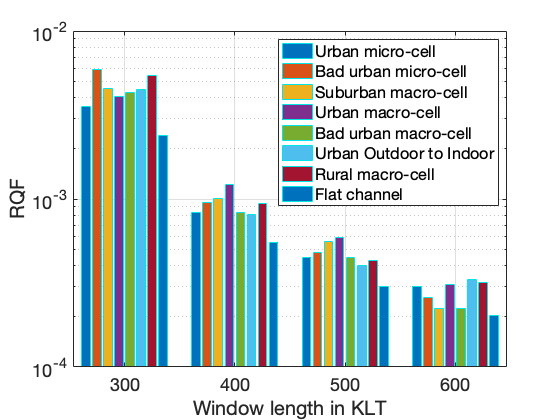}
    \caption{Reconstruction quality estimate over KLT window length - Downlink}
    \label{fig:rqf_dl_winlen}
\end{subfigure}
\begin{subfigure}[b]{0.32\linewidth}
    \includegraphics[width=\linewidth]{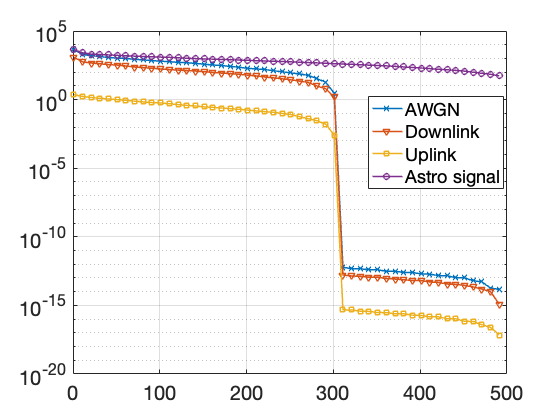}
    \caption{Eigenvalues - Astro, Downlink, Uplink, (AWGN+RFI)}
    \label{fig:eigv_ul_dl}
\end{subfigure}
\caption{Effect of window length on Reconstruction Quality Factor}
\label{fig:uplink_rqf}
\end{figure*}

In \S\ref{sec:collab}, we have emphasized the importance of accurate characterization of RFI for the success of the proposed method. From an empirical estimate, we chose KLT window length $L$ = 500. Figure~\ref{fig:eigv_ul_dl} shows comparison of top 500 eigenvalues from downlink RFI without wireless channel effects (only propagation loss and noise), downlink RFI with channels effects and uplink RFI. As we can observe, both downlink scenarios have comparable eigenvalues. We have to keep in mind that the eigenvalues are not normalized here and thus captures the effect of signal power. This is evident from the eigenvalues of uplink RFI being smaller than those of downlink. Although, they exhibit similar pattern of high eigenvalues up to eigenvalue index 300 followed by a sharp drop beyond. This provides an estimate of the rank of RFI eigenspace. Rest of the eigenvalues correspond to components including noise only. This clearly indicates that accurate characterization of the RFI requires $\sim$300 eigenfunctions as determined by the rank of RFI eigenspace. On contrary, for the astronomical signal, all 500 eigenvalues are comparable. This indicates RFI characterization is possible with window length 300 at BS, but reconstruction quality will degrade for $L{<}$500 and is likely to increase with $L{>}$500. This is because, higher reconstructed signal power is available when the number of residual eigenfunctions are higher after projection and nullifying the RFI subspace of rank 300.
Such hypothesis is supported by the results reported here as well. Figures~\ref{fig:rqf_ul_winlen} and \ref{fig:rqf_dl_winlen} shows the reconstruction quality for different channel conditions along with AWGN channel for window length varying from 300 to 600. These RQFs are measured at a distance of 1000 meters (8-15 dB INR - as noise floor at receiver varies based on modulation and coding rate) from the BS for uplink scenario, as it provides a parity among all the channel conditions discussed here. In both downlink and uplink scenario, there is a steady increase in RQF with decreasing KLT window length. RQF for flat faded channel is less for any given window length than any other channel condition owing to the fact that severe channel deterioration of signal causes poor characterization. Similarly, increase in RQF in uplink compared to downlink is caused by differences in transmit power level in downlink and uplink and propagation loss before characterization at BS.  We observe that for $L{>}$300, there is a sharp improvement in RQF, which eventually moves towards a saturation with growing $L$.

\subsubsection{Impact on real time collaboration}
\label{sec:rate_budget}
The number of dominant eigenvalues or the number of eigenfunctions needed for accurate representation of RFI signals drives the amount of overhead - data that is required to be shared between radio telescope and RFI source. 

\paragraph{Rate budget}
Assuming $M$ eigenfunctions are necessary for characterization of RFI subspace, the eigenfunction matrix will be a complex valued matrix of shape ($L\times M$). Each complex entry will occupy at least 32 bits as a minimum of signed 16 bit depth for both real and imaginary components is necessary to maintain accuracy of shared eigenspace. To match the RFI characterization to that of telescope captured data format, they are channelized accordingly based on the telescope fine channel bandwidth (30.5 kHz) as described in \S\ref{sec:rfi_inject}. This contributes to an eigenspace matrix for each channel leading to a data rate requirement (${r}_d$):
\begin{equation}
    r_d = \frac{1}{\textrm{signal duration}} \times {L} \times {M} \times 32 \times \frac{{B}}{30.5 {kHz}} {bps.}
\end{equation}
Plugging values from our simulation and signal captures, we get:
\begin{equation}
    r_d  = \frac{1}{4} \times 500 \times 300 \times 32 \frac{20 {MHz}}{30.5 {kHz}} {bps.} = 786.67 {Mbps.} \nonumber
\end{equation}

The data rate in reality can be much lower as LTE signal being modulated deterministic signal, its eigenspace does not change significantly across the fine channel bandwidth. Thus, it may suffice to share characterization of LTE across one channel or a few channels instead of all hundreds of them. Additionally, this can be on demand or periodic based on the use case, but even the minimal data rate, assuming one fine channel characterization is shared for the entire bandwidth, remains significant ($\sim$ 1.2 Mbps) for real time processing and storage. This may necessitate the burden of upgrades in software, hardware and storage at telescope and at RFI source. This data-rate is likely to increase with more complicated signal, complex cell topology contributing to the RFI, different traffic pattern and variation with time and distance. To alleviate this issue, the parameter set needs to be more concise and that pushes us to look at non-linear decomposition of signals, generating a set of smaller number non-linear basis functions that are able to accurately characterize the signal. We expect to address such issues in our future work.

%% file: related.tex
\section{Related Work}
\label{sec:related}

\noindent \textbf{RFI mitigation in radio astronomy:}
%
Active RFI mitigation has become a necessary practice in the radio astronomy community.
Known persistent and fixed sources of RFI are highly attenuated at the front-end of the receiver using series of analog superconductive filters \cite{7833543, bolli2012superconducting}, but frequencies with high RFI density (e.g. FM or Digital Video Broadcast band) are usually simply avoided by design. Fast processors (RFSoC and FPGA) enable the detection and the blanking of impulsive RFI in the baseband digital samples, post-digitization \cite{dumez2016multi,ait2010rfi}.
Data flagging consists of detecting time and frequency data corrupted with RFI, and discard them by replacing these values with zeros or random noise. This process occurs after channelization and time integration, where the intermediate telescope data product has the appropriate time resolution (of the order of 1 ms) to match most RFI duty cycles. This can be applied manually after a careful inspection of the collected data, but conventional data reduction software include automated flagger based on local and global statistics of a given dataset \cite{urvashi2003automatic,offringa2015low}. Research in RFI flagging has also resulted in the development of real-time ``on-the-fly'' data flaggers \cite{sclocco2019real}, and the use of machine learning to automatically recognize and classify detected RFI \cite{wolfaardt2016machine}.

Telescope arrays provide spatial information in addition to the time and frequency signatures of the studied objects captured in the telescope correlation matrix. RFI spatial signatures can be extracted from this matrix in order to build an adapted spatial filter, and eventually recover uncorrupted time and frequency data \cite{hellbourg2016spatial}. These methods remain at an experimental level due to their impact on the array calibration. Finally, the subtraction of an estimated and reconstructed RFI waveform from the telescope data have been demonstrated, but have never been deployed due to their heavy computational complexity \cite{ellingson2001removal}.

\noindent \textbf{Interference cancellation in communication systems:} 
Collaboration among wireless technologies~\cite{saha_dyspan_15,survey_spec_infer_18}, 
or avoiding incumbents~\cite{incumbent_19}, 
primarily employ sensing and database management for active users, that cannot be extrapolated to passive users due to an absence of active transmissions. 
Interference cancellation 
in wireless communication~\cite{survey_interf_cancel_13, full_duplex_17},
require decoding the strongest signal first in order to 
cancel it. 
These cannot be applied for RFI mitigation 
or coexistence of active and passive users because:
a) the astronomical signal is not a modulated signal with known characteristics, 
b) the RFI 
at the passive user 
is of extremely low power, which cannot be decoded to remove it, and 
c) sharing active communication signals as digital samples 
with passive users 
for cancellation at the telescope, incurs prohibitive bandwidth. 
Hence, it is essential to accurately characterize RFI in a condensed format 
such that it can be shared with passive users for cancellation, while preserving user privacy and adhering to cellular standards.

\noindent \textbf{Basis expansion and orthogonal projections:} 
Since, KLT incurs high computational complexity to extract a very large number of eigenfunctions \cite{Maccone2012BAMKLT}, we rely on SSA to implement KLT 
which extracts only the $L$ eigenfunctions with largest eigenvalues.
%
Spatial filtering of interference or noise 
projects the signal onto the null space of the undesired components \cite{boonstra2005RFIthesis}. 
Moreover, methods for signal separation \cite{Golyandina2013SSABook}, 
reconstruct the desired signal by identifying correlations 
in the eigenvectors. 
These typically assume 
orthogonality of the subspaces \cite{Golyandina2013SSABook}, 
which does not always hold true in practice. 
Non-orthogonal projections like oblique projection \cite{hellbourg2012oblique} 
reduces the power distortion of the reconstructed signal even when the orthogonality of astronomical and RFI subspaces are not verified, however requires accurate estimation of null-spaces of RFI and column-spaces of astronomical signals which is often challenging when both signals are weak or non-disjoint. 
In contrast, collaborative RFI cancellation does not require the orthogonality of the astronomical and RFI subspaces at the telescope to nullify the RFI, since the RFI subspace is accurately identified at high SNR at the BS.

%% file: conclusion.tex
\section{Conclusion}
\label{sec:conclusion}

This work advances the literature on RFI mitigation for radio astronomy by sharing stochastic characterization of the RFI at its source, the cellular base station, with the telescope to cancel the incident RFI. The method has the potential to be deployed at an actual radio telescope, like DSA-110 at OVRO, and promote collaborative spectrum sharing between the active and passive users of the spectrum. The high reconstruction quality of the RFI-free astronomical signal in our evaluations will further motivate both research  communities to apply the method to eliminate other forms of RFI in various bands allocated for radio astronomy and other passive services. However, managing computational complexity of large eigenvalue problem remain a challenge for real-time RFI cancellation.

%% file: main.bbl
\begin{thebibliography}{10}
\providecommand{\url}[1]{#1}
\csname url@samestyle\endcsname
\providecommand{\newblock}{\relax}
\providecommand{\bibinfo}[2]{#2}
\providecommand{\BIBentrySTDinterwordspacing}{\spaceskip=0pt\relax}
\providecommand{\BIBentryALTinterwordstretchfactor}{4}
\providecommand{\BIBentryALTinterwordspacing}{\spaceskip=\fontdimen2\font plus
\BIBentryALTinterwordstretchfactor\fontdimen3\font minus
  \fontdimen4\font\relax}
\providecommand{\BIBforeignlanguage}[2]{{%
\expandafter\ifx\csname l@#1\endcsname\relax
\typeout{** WARNING: IEEEtran.bst: No hyphenation pattern has been}%
\typeout{** loaded for the language `#1'. Using the pattern for}%
\typeout{** the default language instead.}%
\else
\language=\csname l@#1\endcsname
\fi
#2}}
\providecommand{\BIBdecl}{\relax}
\BIBdecl

\bibitem{quiet_skies_20}
\BIBentryALTinterwordspacing
I.~A. Organisation, \emph{Dark and Quiet Skies for Science and Society}, 2020.
  [Online]. Available:
  \url{https://www.iau.org/static/publications/dqskies-book-29-12-20.pdf}
\BIBentrySTDinterwordspacing

\bibitem{nap21729_15}
\BIBentryALTinterwordspacing
{National Academies of Sciences, Engineering, and Medicine}, \emph{A Strategy
  for Active Remote Sensing Amid Increased Demand for Radio Spectrum}.\hskip
  1em plus 0.5em minus 0.4em\relax Washington, DC: The National Academies
  Press, 2015. [Online]. Available:
  \url{https://www.nap.edu/catalog/21729/a-strategy-for-active-remote-sensing-amid-increased-demand-for-radio-spectrum}
\BIBentrySTDinterwordspacing

\bibitem{fcc_rural}
\BIBentryALTinterwordspacing
{Federal Communication Commission}, ``{2020 FCC Broadband Deployment Plan},''
  \emph{2020 BROADBAND DEPLOYMENT REPORT}, vol.~21, no.~1, pp. 1--9, 2020.
  [Online]. Available:
  \url{https://www.fcc.gov/reports-research/reports/broadband-progress-reports/2020-broadband-deployment-report}
\BIBentrySTDinterwordspacing

\bibitem{rau2019rfi}
U.~Rau, R.~Selina, and A.~Erickson, ``Rfi mitigation for the ngvla: A
  cost-benefit analysis ngvla memo\# 70,'' 2019.

\bibitem{8742126}
J.~E. Velazco, L.~Ledezma, J.~Bowen, L.~Samoska, M.~Soriano, A.~Akgiray,
  S.~Weinreb, and J.~Lazio, ``Ultra-wideband low noise amplifiers for the next
  generation very large array,'' in \emph{2019 IEEE Aerospace Conference},
  2019, pp. 1--6.

\bibitem{9440921}
J.~Shi and S.~Weinreb, ``Temperature compensated internal lna noise calibration
  source,'' \emph{IEEE Microwave and Wireless Components Letters}, vol.~31,
  no.~8, pp. 1016--1019, 2021.

\bibitem{8879003}
K.~Jeganathan, A.~Dunning, Y.~S. Chung, M.~Bourne, M.~Bowen, S.~Castillo,
  N.~Carter, P.~Doherty, D.~George, D.~Hayman, S.~Mackay, L.~Reilly,
  P.~Roberts, P.~Roush, S.~Severs, K.~Smart, R.~Shaw, S.~Smith, and J.~Tuthill,
  ``Ultra wideband (uwb) receiver for radio astronomy,'' in \emph{2019
  International Conference on Electromagnetics in Advanced Applications
  (ICEAA)}, 2019, pp. 0343--0346.

\bibitem{national2007handbook}
N.~R. Council \emph{et~al.}, \emph{Handbook of Frequency Allocations and
  Spectrum Protection for Scientific Uses}.\hskip 1em plus 0.5em minus
  0.4em\relax National Academies Press, 2007.

\bibitem{hallinan2019dsa}
G.~Hallinan, V.~Ravi, S.~Weinreb, J.~Kocz, Y.~Huang, D.~Woody, J.~Lamb,
  L.~D'Addario, M.~Catha, J.~Shi \emph{et~al.}, ``The dsa-2000--a radio survey
  camera,'' \emph{arXiv preprint arXiv:1907.07648}, 2019.

\bibitem{Zaroubi_2012}
\BIBentryALTinterwordspacing
S.~Zaroubi, ``The epoch of reionization,'' in \emph{The First Galaxies}.\hskip
  1em plus 0.5em minus 0.4em\relax Springer Berlin Heidelberg, sep 2012, pp.
  45--101. [Online]. Available:
  \url{https://doi.org/10.1007%2F978-3-642-32362-1_2}
\BIBentrySTDinterwordspacing

\bibitem{NAP21774}
\BIBentryALTinterwordspacing
{{National Academies of Sciences Engineering and Medicine}}, \emph{Handbook of
  Frequency Allocations and Spectrum Protection for Scientific Uses: Second
  Edition}.\hskip 1em plus 0.5em minus 0.4em\relax Washington, DC: The National
  Academies Press, 2015. [Online]. Available:
  \url{https://www.nap.edu/catalog/21774/handbook-of-frequency-allocations-and-spectrum-protection-for-scientific-uses}
\BIBentrySTDinterwordspacing

\bibitem{itu_radio_astronomy_13}
{{The International Telecommunication Union - Radiocommunication Bureau}},
  \emph{{{Handbook on Radio Astronomy}}}.\hskip 1em plus 0.5em minus
  0.4em\relax {{The International Telecommunication Union - Radiocommunication
  Bureau}}, 2013.

\bibitem{craf_radio_astronomy_05}
\BIBentryALTinterwordspacing
{{Committee on Radio Astronomy Frequencies (CRAF)}}, \emph{{{CRAF Handbook for
  Radio Astronomy}}}.\hskip 1em plus 0.5em minus 0.4em\relax EUROPEAN SCIENCE
  FOUNDATION, 2005. [Online]. Available:
  \url{https://craf.eu/wp-content/uploads/2015/02/CRAFhandbook3.pdf}
\BIBentrySTDinterwordspacing

\bibitem{pawr-rural}
``{National Science Foundation-Funded Wireless Testbed on Rural Broadband},''
  \url{https://advancedwireless.org/national-science-foundation-funded-pawr-program-selects-two-finalists-for-fourth-wireless-testbed/}.

\bibitem{kocz2019dsa}
J.~Kocz, V.~Ravi, M.~Catha, L.~D’Addario, G.~Hallinan, R.~Hobbs, S.~Kulkarni,
  J.~Shi, H.~Vedantham, S.~Weinreb \emph{et~al.}, ``Dsa-10: a prototype array
  for localizing fast radio bursts,'' \emph{Monthly Notices of the Royal
  Astronomical Society}, vol. 489, no.~1, pp. 919--927, 2019.

\bibitem{rfi_dyspan_2021}
M.~Careem, S.~Chakaraborty, A.~Dutta, D.~Saha, and G.~Hellbourg, ``Spectrum
  sharing via collaborative rfi cancellation for radio astronomy,'' in
  \emph{2021 IEEE International Symposium on Dynamic Spectrum Access Networks
  (DySPAN)}, 2021, pp. 97--104.

\bibitem{wilson2009tools}
T.~L. Wilson, K.~Rohlfs, and S.~H{\"u}ttemeister, \emph{Tools of radio
  astronomy}.\hskip 1em plus 0.5em minus 0.4em\relax Springer, 2009, vol.~5.

\bibitem{taylor1999synthesis}
G.~B. Taylor, C.~L. Carilli, and R.~A. Perley, ``Synthesis imaging in radio
  astronomy ii,'' \emph{Synthesis Imaging in Radio Astronomy II}, vol. 180,
  1999.

\bibitem{price2021spectrometers}
D.~C. Price, ``Spectrometers and polyphase filterbanks in radio astronomy,'' in
  \emph{The WSPC Handbook of Astronomical Instrumentation: Volume 1: Radio
  Astronomical Instrumentation}.\hskip 1em plus 0.5em minus 0.4em\relax World
  Scientific, 2021, pp. 159--179.

\bibitem{bem_book}
\BIBentryALTinterwordspacing
T.~Hastie, J.~Friedman, and R.~Tibshirani, \emph{Basis Expansions and
  Regularization}.\hskip 1em plus 0.5em minus 0.4em\relax New York, NY:
  Springer New York, 2001, pp. 115--163. [Online]. Available:
  \url{https://doi.org/10.1007/978-0-387-21606-5_5}
\BIBentrySTDinterwordspacing

\bibitem{RFI_KLT}
A.~Szumski and G.~Hein, ``Finding the interference karhunen-lo{\`e}ve transform
  as an instrument to detect weak rf signals,'' in \emph{InsideGNSS}, 2011, p.
  57–64.

\bibitem{trudu2020performance}
M.~Trudu, M.~Pilia, G.~Hellbourg, P.~Pari, N.~Antonietti, C.~Maccone, A.~Melis,
  D.~Perrodin, and A.~Trois, ``Performance analysis of the karhunen--lo{\`e}ve
  transform for artificial and astrophysical transmissions: denoizing and
  detection,'' \emph{Monthly Notices of the Royal Astronomical Society}, vol.
  494, no.~1, pp. 69--83, 2020.

\bibitem{tome2018use}
A.~M. Tomé, D.~Malafaia, A.~R. Teixeira, and E.~W. Lang, ``On the use of
  singular spectrum analysis,'' 2018.

\bibitem{ssa}
\BIBentryALTinterwordspacing
R.~Vautard, P.~Yiou, and M.~Ghil, ``Singular-spectrum analysis: A toolkit for
  short, noisy chaotic signals,'' \emph{Physica D: Nonlinear Phenomena},
  vol.~58, no.~1, pp. 95--126, 1992. [Online]. Available:
  \url{https://www.sciencedirect.com/science/article/pii/016727899290103T}
\BIBentrySTDinterwordspacing

\bibitem{Szumski2011KLT}
A.~Szumski, ``Karhunen loeve transform as an instrument to detect weak rf
  signals,'' InsideGNSS, European Space Agency, Tech. Rep., 2011.

\bibitem{Golyandina2013SSABook}
N.~Golyandina and A.~Zhigljavsky, \emph{Singular Spectrum Analysis for Time
  Series}, 01 2013.

\bibitem{raprotection}
R.~I.-R. RA.769-2, ``Protection criteria used for radio astronomical
  measurements.''

\bibitem{ovro}
\BIBentryALTinterwordspacing
 [Online]. Available: \url{https://www.ovro.caltech.edu/}
\BIBentrySTDinterwordspacing

\bibitem{petroff2019fast}
E.~Petroff, J.~Hessels, and D.~Lorimer, ``Fast radio bursts,'' \emph{The
  Astronomy and Astrophysics Review}, vol.~27, no.~1, pp. 1--75, 2019.

\bibitem{walsh2018searching}
G.~Walsh and R.~Lynch, ``Searching for single pulses using heimdall,'' in
  \emph{American Astronomical Society Meeting Abstracts\# 231}, vol. 231, 2018,
  pp. 243--04.

\bibitem{van2021psrdada}
W.~van Straten, A.~Jameson, and S.~Os{\l}owski, ``Psrdada: Distributed
  acquisition and data analysis for radio astronomy,'' \emph{Astrophysics
  Source Code Library}, pp. ascl--2110, 2021.

\bibitem{dickey1990hi}
J.~M. Dickey and F.~J. Lockman, ``Hi in the galaxy,'' \emph{Annual review of
  astronomy and astrophysics}, vol.~28, pp. 215--261, 1990.

\bibitem{winner2_channel}
P.~Kyösti, J.~Meinilä, L.~Hentila, X.~Zhao, T.~Jämsä, C.~Schneider,
  M.~Narandzi'c, M.~Milojevi'c, A.~Hong, J.~Ylitalo, V.-M. Holappa,
  M.~Alatossava, R.~Bultitude, Y.~Jong, and T.~Rautiainen, ``Ist-4-027756
  winner ii d1.1.2 v1.2 winner ii channel models,'' \emph{Inf. Soc. Technol},
  vol.~11, 02 2008.

\bibitem{rqf_ssa}
\BIBentryALTinterwordspacing
J.~Harmouche, D.~Fourer, F.~Auger, P.~Borgnat, and P.~Flandrin, ``{The Sliding
  Singular Spectrum Analysis: a Data-Driven Non-Stationary Signal Decomposition
  Tool},'' \emph{{IEEE Transactions on Signal Processing}}, Sep. 2017.
  [Online]. Available: \url{https://hal.archives-ouvertes.fr/hal-01589464}
\BIBentrySTDinterwordspacing

\bibitem{boonstra2005RFIthesis}
A.-J. Boonstra, ``Radio frequency interference mitigation in radio astronomy,''
  Ph.D. dissertation, Technische Universiteit Delft, 2005.

\bibitem{GPSgovGP35:online}
``Gps.gov: Gps accuracy,''
  \url{https://www.gps.gov/systems/gps/performance/accuracy/#timing}, (Accessed
  on 10/15/2021).

\bibitem{7833543}
A.~Soliman, S.~Weinreb, G.~Rajagopalan, C.~Eckert, and L.~Hilliard,
  ``Quadruple-ridged flared horn feed with internal rfi band rejection
  filter,'' in \emph{2016 Radio Frequency Interference (RFI)}, 2016, pp.
  115--116.

\bibitem{bolli2012superconducting}
P.~Bolli and F.~Huang, ``Superconducting filter for radio astronomy using
  interdigitated, capacitively loaded spirals,'' \emph{Experimental Astronomy},
  vol.~33, no.~1, pp. 225--236, 2012.

\bibitem{dumez2016multi}
C.~Dumez-Viou, R.~Weber, and P.~Ravier, ``Multi-level pre-correlation rfi
  flagging for real-time implementation on uniboard,'' \emph{Journal of
  Astronomical Instrumentation}, vol.~5, no.~04, p. 1641019, 2016.

\bibitem{ait2010rfi}
D.~Ait-Allal, C.~Dumez-Viou, R.~Weber, G.~Desvignes, I.~Cognard, and
  G.~Theureau, ``Rfi mitigation at nanc{\c{}} ay observatory: Impulsive signal
  processing,'' in \emph{Widefield Science and Technology for the SKA SKADS
  Conference 2009}.\hskip 1em plus 0.5em minus 0.4em\relax ISBN
  978-90-805434-5-4, 2010, pp. 201--205.

\bibitem{urvashi2003automatic}
R.~Urvashi, A.~P. Rao, and N.~Pune, ``Automatic rfi identification and
  flagging,'' \emph{The National Centre for Radio Astrophysics of the Tata
  Institute of Fundamental Research}, 2003.

\bibitem{offringa2015low}
A.~Offringa, R.~Wayth, N.~Hurley-Walker, D.~Kaplan, N.~Barry, A.~Beardsley,
  M.~Bell, G.~Bernardi, J.~Bowman, F.~Briggs \emph{et~al.}, ``The low-frequency
  environment of the murchison widefield array: radio-frequency interference
  analysis and mitigation,'' \emph{Publications of the Astronomical Society of
  Australia}, vol.~32, 2015.

\bibitem{sclocco2019real}
A.~Sclocco, D.~Vohl, and R.~V. van Nieuwpoort, ``Real-time rfi mitigation for
  the apertif radio transient system,'' in \emph{2019 RFI Workshop-Coexisting
  with Radio Frequency Interference (RFI)}.\hskip 1em plus 0.5em minus
  0.4em\relax IEEE, 2019, pp. 1--8.

\bibitem{wolfaardt2016machine}
C.~J. Wolfaardt, ``Machine learning approach to radio frequency interference
  (rfi) classification in radio astronomy,'' Ph.D. dissertation, Stellenbosch:
  Stellenbosch University, 2016.

\bibitem{hellbourg2016spatial}
G.~Hellbourg, K.~Bannister, and A.~HotarP, ``Spatial filtering experiment with
  the askap beta array,'' in \emph{2016 Radio Frequency Interference
  (RFI)}.\hskip 1em plus 0.5em minus 0.4em\relax IEEE, 2016, pp. 37--42.

\bibitem{ellingson2001removal}
S.~W. Ellingson, J.~D. Bunton, and J.~F. Bell, ``Removal of the glonass c/a
  signal from oh spectral line observations using a parametric modeling
  technique,'' \emph{The Astrophysical Journal Supplement Series}, vol. 135,
  no.~1, p.~87, 2001.

\bibitem{saha_dyspan_15}
S.~{Sagari}, S.~{Baysting}, D.~{Saha}, I.~{Seskar}, W.~{Trappe}, and
  D.~{Raychaudhuri}, ``Coordinated dynamic spectrum management of lte-u and
  wi-fi networks,'' in \emph{2015 IEEE International Symposium on Dynamic
  Spectrum Access Networks (DySPAN)}, 2015, pp. 209--220.

\bibitem{survey_spec_infer_18}
G.~{Ding}, Y.~{Jiao}, J.~{Wang}, Y.~{Zou}, Q.~{Wu}, Y.~{Yao}, and L.~{Hanzo},
  ``Spectrum inference in cognitive radio networks: Algorithms and
  applications,'' \emph{IEEE Communications Surveys Tutorials}, vol.~20, no.~1,
  pp. 150--182, 2018.

\bibitem{incumbent_19}
M.~R. {Souryal} and T.~T. {Nguyen}, ``Effect of federal incumbent activity on
  cbrs commercial service,'' in \emph{2019 IEEE International Symposium on
  Dynamic Spectrum Access Networks (DySPAN)}, 2019, pp. 1--5.

\bibitem{survey_interf_cancel_13}
N.~I. {Miridakis} and D.~D. {Vergados}, ``A survey on the successive
  interference cancellation performance for single-antenna and multiple-antenna
  ofdm systems,'' \emph{IEEE Communications Surveys Tutorials}, vol.~15, no.~1,
  pp. 312--335, 2013.

\bibitem{full_duplex_17}
M.~{Amjad}, F.~{Akhtar}, M.~H. {Rehmani}, M.~{Reisslein}, and T.~{Umer},
  ``Full-duplex communication in cognitive radio networks: A survey,''
  \emph{IEEE Communications Surveys Tutorials}, vol.~19, no.~4, pp. 2158--2191,
  2017.

\bibitem{Maccone2012BAMKLT}
C.~Maccone, \emph{A simple introduction to the KLT and BAM-KLT}, 08 2012, pp.
  411--448.

\bibitem{hellbourg2012oblique}
G.~Hellbourg, R.~Weber, C.~Capdessus, and A.-J. Boonstra, ``Oblique projection
  beamforming for rfi mitigation in radio astronomy,'' in \emph{2012 IEEE
  Statistical Signal Processing Workshop (SSP)}.\hskip 1em plus 0.5em minus
  0.4em\relax IEEE, 2012, pp. 93--96.

\end{thebibliography}
